\documentclass[twocolumn]{aastex631}

\usepackage{amsmath}
\usepackage{multirow}
\usepackage{longtable}

\newcommand {\CL}{A2744}
\newcommand {\LT}{\texttt{LensTool}}
\newcommand {\EM}{$\mathrm{EM}$}
\newcommand {\SLOT}{\texttt{SLOT}}
\newcommand {\T}{Table~}

\newcommand {\Sec}{Section~}
\newcommand {\Fig}{Figure~}
\newcommand {\Eq}{Eq.~}

\defcitealias{Bergamini_2021}{B21}
\defcitealias{Bergamini2022}{B23}
\defcitealias{Richard_2021}{R21}
\defcitealias{Furtak2022}{F22}

\shorttitle{Extended strong lensing modeling of Abell 2744}
\shortauthors{Bergamini et al.}

\begin{document}


\title{The GLASS-JWST Early Release Science Program. III. Strong lensing model of Abell 2744 and its infalling regions}

\correspondingauthor{Pietro Bergamini}
\email{pietro.bergamini@unimi.it}

\author[0000-0003-1383-9414]{Pietro Bergamini}
\affiliation{Dipartimento di Fisica, Universit\`a degli Studi di Milano, Via Celoria 16, I-20133 Milano, Italy}
\affiliation{INAF -- OAS, Osservatorio di Astrofisica e Scienza dello Spazio di Bologna, via Gobetti 93/3, I-40129 Bologna, Italy}

\author[0000-0003-3108-9039]{Ana Acebron}
\affiliation{Dipartimento di Fisica, Universit\`a degli Studi di Milano, Via Celoria 16, I-20133 Milano, Italy}
\affiliation{INAF -- IASF Milano, via A. Corti 12, I-20133 Milano, Italy}

\author[0000-0002-5926-7143]{Claudio Grillo}
\affiliation{Dipartimento di Fisica, Universit\`a degli Studi di Milano, Via Celoria 16, I-20133 Milano, Italy}
\affiliation{INAF -- IASF Milano, via A. Corti 12, I-20133 Milano, Italy}

\author[0000-0002-6813-0632]{Piero Rosati}
\affiliation{Dipartimento di Fisica e Scienze della Terra, Universit\`a degli Studi di Ferrara, via Saragat 1, I-44122 Ferrara, Italy}
\affiliation{INAF -- OAS, Osservatorio di Astrofisica e Scienza dello Spazio di Bologna, via Gobetti 93/3, I-40129 Bologna, Italy}

\author[0000-0001-6052-3274]{Gabriel Bartosch Caminha}
\affiliation{Technische Universit\"at M\"unchen, Physik-Department, James-Franck Str. 1, 85741 Garching, Germany}
\affiliation{Max-Planck-Institut f\"ur Astrophysik, Karl-Schwarzschild-Str. 1, D-85748 Garching, Germany}

\author[0000-0001-9261-7849]{Amata Mercurio}
\affiliation{Dipartimento di Fisica “E.R. Caianiello”, Università Degli Studi di Salerno, Via Giovanni Paolo II, I–84084 Fisciano (SA), Italy}
\affiliation{INAF -- Osservatorio Astronomico di Capodimonte, Via Moiariello 16, I-80131 Napoli, Italy}

\author[0000-0002-5057-135X]{Eros Vanzella}
\affiliation{INAF -- OAS, Osservatorio di Astrofisica e Scienza dello Spazio di Bologna, via Gobetti 93/3, I-40129 Bologna, Italy}

\author[0000-0002-3407-1785]{Charlotte Mason}
\affiliation{Cosmic Dawn Center (DAWN), Jagtvej 128, DK2200 Copenhagen N, Denmark}
\affiliation{Niels Bohr Institute, University of Copenhagen, Jagtvej 128, København N, DK-2200, Denmark}

\author[0000-0002-8460-0390]{Tommaso Treu}
\affiliation{Department of Physics and Astronomy, University of California, Los Angeles, 430 Portola Plaza, Los Angeles, CA 90095, USA}

\author[0000-0002-0316-6562]{Giuseppe Angora}
\affiliation{Dipartimento di Fisica e Scienze della Terra, Universit\`a degli Studi di Ferrara, via Saragat 1, I-44122 Ferrara, Italy}
\affiliation{INAF -- Osservatorio Astronomico di Capodimonte, Via Moiariello 16, I-80131 Napoli, Italy}

\author[0000-0003-2680-005X]{Gabriel B. Brammer}
\affiliation{Cosmic Dawn Center (DAWN), Jagtvej 128, DK2200 Copenhagen N, Denmark}
\affiliation{Niels Bohr Institute, University of Copenhagen, Jagtvej 128, København N, DK-2200, Denmark}

\author[0000-0003-1225-7084]{Massimo Meneghetti}
\affiliation{INAF -- OAS, Osservatorio di Astrofisica e Scienza dello Spazio di Bologna, via Gobetti 93/3, I-40129 Bologna, Italy}
\affiliation{National Institute for Nuclear Physics, viale Berti Pichat 6/2, I-40127 Bologna, Italy}

\author[0000-0001-6342-9662]{Mario Nonino}
\affiliation{INAF -- Osservatorio Astronomico di Trieste, via G. B. Tiepolo 11, I-34131 Trieste, Italy}

\author[0000-0003-4109-304X]{Kristan Boyett}
\affiliation{School of Physics, University of Melbourne, Parkville 3010, VIC, Australia}
\affiliation{ARC Centre of Excellence for All Sky Astrophysics in 3 Dimensions (ASTRO 3D), Australia} 

\author[0000-0001-5984-0395]{Maru\v{s}a Brada\v{c}}
\affiliation{University of Ljubljana, Department of Mathematics and Physics, Jadranska ulica 19, SI-1000 Ljubljana, Slovenia} 
\affiliation{Department of Physics and Astronomy, University of California Davis, 1 Shields Avenue, Davis, CA 95616, USA} 

\author[0000-0001-9875-8263]{Marco Castellano}
\affiliation{INAF -- Osservatorio Astronomico di Roma, Via Frascati 33, 00078 Monteporzio Catone, Rome, Italy}

\author[0000-0003-3820-2823]{Adriano Fontana}
\affiliation{INAF -- Osservatorio Astronomico di Roma, Via Frascati 33, 00078 Monteporzio Catone, Rome, Italy}

\author[0000-0002-8512-1404]{Takahiro Morishita}
\affiliation{IPAC, California Institute of Technology, MC 314-6, 1200 E. California Boulevard, Pasadena, CA 91125, USA}

\author[0000-0002-7409-8114]{Diego Paris}
\affiliation{INAF -- Osservatorio Astronomico di Roma, Via Frascati 33, 00078 Monteporzio Catone, Rome, Italy}

\author[0000-0003-3518-0374]{Gonzalo Prieto-Lyon}
\affiliation{Cosmic Dawn Center (DAWN), Jagtvej 128, DK2200 Copenhagen N, Denmark}
\affiliation{Niels Bohr Institute, University of Copenhagen, Jagtvej 128, København N, DK-2200, Denmark}

\author[0000-0002-4140-1367]{Guido Roberts-Borsani}
\affiliation{Department of Physics and Astronomy, University of California, Los Angeles, 430 Portola Plaza, Los Angeles, CA 90095, USA}

\author[0000-0002-4430-8846]{Namrata Roy}
\affiliation{Center for Astrophysical Sciences, Department of Physics \& Astronomy, Johns Hopkins University, Baltimore, MD 21218, USA.}

\author[0000-0002-9334-8705]{Paola Santini}
\affiliation{INAF -- Osservatorio Astronomico di Roma, Via Frascati 33, 00078 Monteporzio Catone, Rome, Italy}

\author[0000-0003-0980-1499]{Benedetta Vulcani}
\affiliation{INAF -- Osservatorio astronomico di Padova, Vicolo Osservatorio 5, I-35122 Padova, Italy} 

\author[0000-0002-9373-3865]{Xin Wang}
\affil{School of Astronomy and Space Science, University of Chinese Academy of Sciences (UCAS), Beijing 100049, China}
\affil{National Astronomical Observatories, Chinese Academy of Sciences, Beijing 100101, China}
\affil{Institute for Frontiers in Astronomy and Astrophysics, Beijing Normal University,  Beijing 102206, China}

\author[0000-0002-8434-880X]{Lilan Yang}
\affiliation{Kavli Institute for the Physics and Mathematics of the Universe, The University of Tokyo, Kashiwa, Japan 277-8583}



\begin{abstract}
We present a new high-precision, JWST-based, strong lensing model for the galaxy cluster Abell 2744 at $z=0.3072$. 
By combining the deep, high-resolution JWST imaging from the GLASS-JWST and UNCOVER programs and a Director's Discretionary Time program, with newly obtained VLT/MUSE data, we identify 32 multiple images from 11 background sources lensed by two external sub-clusters at distances of $\sim160\arcsec$ from the main cluster. The new MUSE observations enable the first spectroscopic confirmation of a multiple image system in the external clumps.  
Moreover, the re-analysis of the spectro-photometric archival and JWST data yields 27 additional multiple images in the main cluster. 
The new lens model is constrained by 149 multiple images ($\sim66\%$ more than in our previous \citealt{Bergamini2022} model) covering an extended redshift range between 1.03 and 9.76. The subhalo mass component of the cluster includes 177 member galaxies down to $m_{\rm F160W}=21$, 163 of which are spectroscopically confirmed. Internal velocity dispersions are measured for 85 members. 
The new lens model is characterized by a remarkably low scatter between predicted and observed positions of the multiple images ($0.43\arcsec$). This precision is unprecedented given the large multiple image sample, the complexity of the cluster mass distribution, and the large modeled area. 
The improved accuracy and resolution of the cluster total mass distribution provides a robust magnification map over a $\sim\!45$ arcmin$^2$ area, which is critical for inferring the intrinsic physical properties of the highly magnified, high-$z$ sources.  
The lens model and the new MUSE redshift catalog are released with this publication.
\end{abstract}

\keywords{Galaxy Cluster (584) ---  Strong Gravitational Lensing (1643) ---  Dark Matter (353)}


\section{Introduction}

The first JWST observations of massive galaxy clusters have revealed exceptionally rich background patterns of strongly lensed galaxies, thanks to a revolutionary combination of angular resolution, depth and near-and mid-infrared coverage \citep[e.g.,][]{Mahler2022, Pascale22, Caminha22}. Several JWST survey programs have targeted cluster fields to exploit the gravitational lensing magnification, in the effort to unveil primordial star-forming systems and study their physical properties \citep[e.g.,][]{Treu2022, Bezanson2022, Willott2022, Windhorst2023}. While current studies are identifying an increasing number of  strongly lensed sources at $z\gtrsim 8$ based on JWST/NIRCam photometry \citep{Hsiao2022,Adams2023,Bradley2022}, some have already obtained spectroscopic confirmations in the range $z=9.5-10$ \citep{Roberts-Borsani2022a, Williams2022}.

Compared to blank field surveys, observations of highly magnified regions of clusters have the advantage of extending to low luminosities and stellar masses our knowledge of the galaxy populations in the first billion years of cosmic history, providing unique insights on their inner structure down to parsec resolutions \citep[e.g.,][]{Bouwens21,Vanzella2022,Mestric2022, Vanzella2023, Welch2023}. 
This progress, however, comes with significant challenges. High-precision strong lensing models are needed to produce reliable magnification maps which are, in turn, critical to infer the intrinsic physical properties of the lensed sources (luminosities, stellar masses, star formation rates, sizes), as well as the effective survey volume at varying redshifts and therefore their space densities \citep{Castellano2022}. The accuracy and precision of lens models in predicting accurate magnification ($\mu$) values depend essentially on the number of bona-fide multiple images with spectroscopic redshifts, spanning a wide $z$-range, and a complete knowledge of the member galaxies contributing to the cluster mass distribution. While independent models based on high-quality spectro-photometric data tend to be robust in the low-magnification regime \citep[$\mu\lesssim 3$, see e.g.,][]{Meneghetti_2017}, such a requirement becomes progressively more important near the critical lines, at $\mu\gtrsim 10$. In the most extreme cases, for sources found within a fraction of arcseconds of a critical line \citep[e.g.,][]{Chen2022, Diego2022, Welch2022, Meena2022}, including caustic crossing events \citep{Kelly2018}, magnification values can be of the order of hundreds or more, implying that the interpretation on the nature of the source relies heavily on the lens model accuracy. 

JWST observations of galaxy clusters with exceptional lensing cross sections offer an unprecedented opportunity to zoom into primordial star-forming regions. On the one hand, they also provide a high number density of multiple image systems, a factor of 2-3 higher than HST based studies, which improves the accuracy of the lens models. On the other hand, spectroscopic redshifts of a significant fraction of these lensed sources, that were beyond the reach of previous technology, have now become attainable with JWST.

In this context, the massive galaxy cluster Abell 2744 (\CL\ hereafer), at $z=0.3072$, has been the target of several observational campaigns with the Hubble Space Telescope (HST) obtaining deep, high-resolution and wide-field coverage \citep{Lotz_2017HFF, Steinhardt2020}. Its combination with extensive ground and space based spectroscopic follow-up observations \citep{Braglia2009,Owers2011, Treu_2015, Richard2021} has enabled a new generation of strong lensing models including a large number of secure multiple images \citep[][\citetalias{Bergamini2022} hereafter]{Wang.2015, Richard2021, Bergamini2022}. In particular, the lens model presented by \citetalias{Bergamini2022} prior to the JWST observations included 90 spectroscopically confirmed multiple images (from 30 background sources), achieving an accuracy in reproducing the observed positions of the multiple images a factor of $2$ better than previous lens models \cite[see e.g.,][]{Wang.2015, Mahler_2018, Richard2021}. 
These pre-JWST lens models found that several massive structures residing at large distances from the main cluster core had a non-negligible impact in the total mass reconstruction of the cluster inner regions. \citetalias{Bergamini2022} found magnification to be significantly different from one well beyond the innermost 500 kpc from the cluster center, suggesting that these specific regions cannot effectively be considered as blank fields.

\begin{figure*}
        \centering
        \includegraphics[width=1\linewidth]{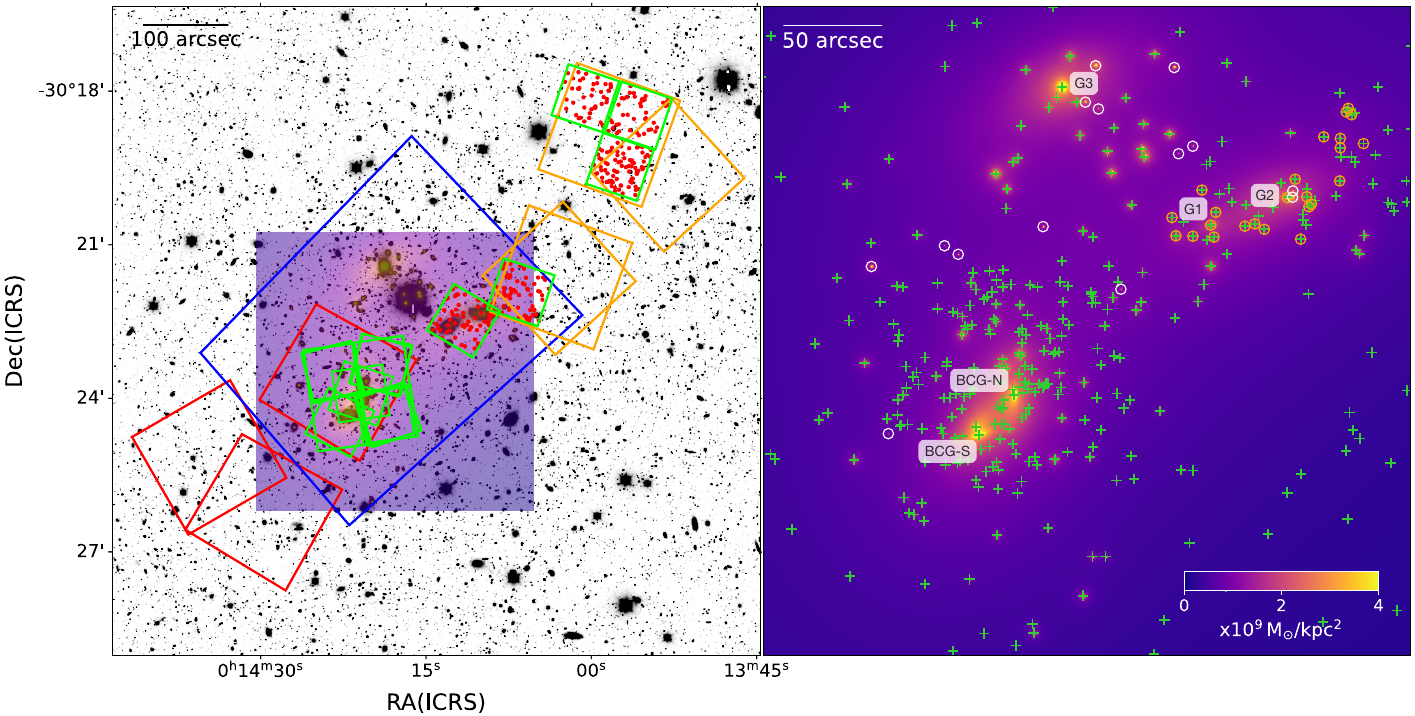}
        \caption{JWST/NIRCam and VLT/MUSE pointings observed for the galaxy cluster \CL\ and the projected total mass density distribution obtained from the extended lens model presented in this work. \textit{Left:} Magellan $r$-band image of the galaxy cluster \CL. The footprints of the NIRCam pointings observed during the GLASS-JWST ERS, UNCOVER, and DTT-2756 programs are shown in orange, blue, and red, respectively. The footprints of the MUSE pointings are drawn in green. The magenta squared region represents the $6.7\times 6.7$ arcmin$^2$ area covered by the extended lens model. Red dots show the 313 newly determined secure redshifts from the MUSE DDT observations. \textit{Right:} Projected total mass density distribution of \CL\ obtained from the extended lens model. The green crosses mark the position of the spectroscopically confirmed cluster member galaxies, corresponding to the light-gray histogram in \Fig \ref{fig:hist_CM}. We note that 163 spectroscopic cluster galaxies with $m_\mathrm{F160W}<21$, plus 14 non-spectroscopically confirmed galaxies (white circles), are included in the lens model. The 24 cluster galaxies included in the lens model with a new spectroscopic confirmation from the MUSE DDT data are encircled in orange.
        }
        \label{fig:TotClusterCM}
\end{figure*}

\begin{figure}[ht!]
        \centering
        \includegraphics[width=1\linewidth]{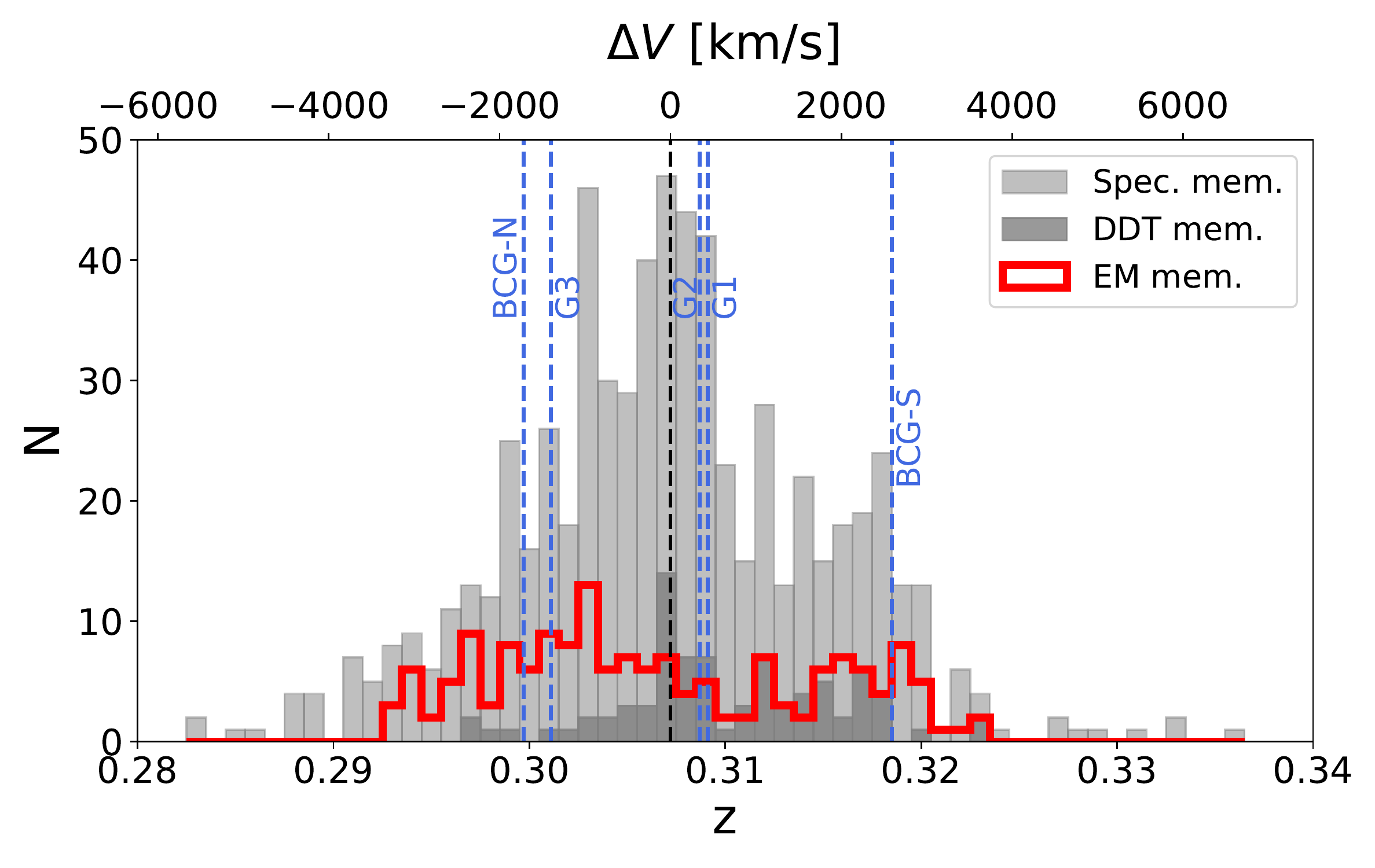}
        \caption{Redshift distribution of the spectroscopic cluster member galaxies of \CL. The redshift distribution of all the  spectroscopically confirmed cluster galaxies within the redshift interval $0.28 \le z \le 0.34$ is plotted using the light gray histogram (669 galaxies). The dark gray histogram corresponds to the distribution of the newly identified members, with secure redshift values, measured from the DDT MUSE data (82 galaxies). The red histogram shows the redshift distribution of the spectroscopically confirmed cluster member galaxies brighter than $m_{\mathrm{F160W}}=21$ considered in the lens model (163 galaxies, see also \Fig \ref{fig:TotClusterCM}).}
        \label{fig:hist_CM}
\end{figure}

\begin{figure}[ht!]
        \centering
        \includegraphics[width=1\linewidth]{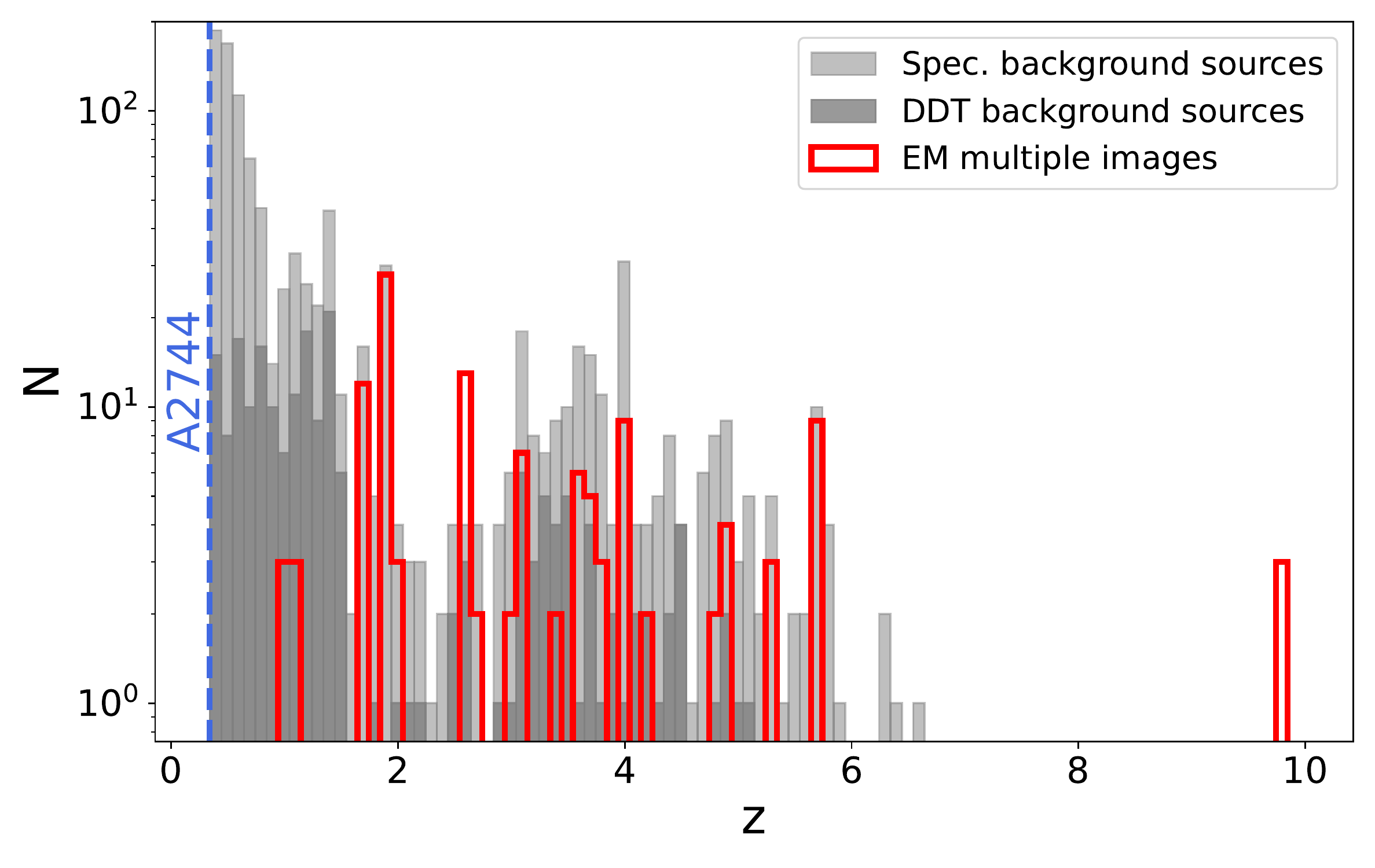}
        \caption{Redshift distribution of all the sources with a secure spectroscopic redshift higher than the \CL\ cluster redshift. The redshift distribution of all the galaxies with $z>0.34$ is plotted using the light gray histogram. The dark gray histogram corresponds to the distribution of the background sources with secure redshift values measured from the new DDT MUSE data. The red histogram shows the redshift distribution of the spectroscopically confirmed multiple images considered in the lens model. All of the 121 spectroscopic multiple images, including the lensed galaxy substructures, are included in the plot (see also \Fig \ref{fig:TotClusterImage}).}
        \label{fig:hist_IM}
\end{figure}

More recently, \CL\ has been among the first targets of JWST \citep{Treu2022, Bezanson2022}.
By exploiting the deep JWST/NIRCam observations gathered from both programs, \citealt{Furtak2022} (\citetalias{Furtak2022} hereafter) have presented a lens model, identifying, for the first time, a large number of photometric multiple images around the external clumps. The \citetalias{Furtak2022} lens model (optimized on the source plane) is constrained by the observed positions of 138 multiple images (from 48 background sources) and it is characterized by an r.m.s residual scatter on the lens plane of  $0.66\arcsec$ in the multiple image positions, compared to $0.37\arcsec$ for the model by \citetalias{Bergamini2022}. 

In this work, we exploit the new JWST/NIRCam imaging and Multi Unit Spectroscopic Explorer (MUSE) spectroscopic observations, together with archival spectro-photometric data including recent JWST/NIRISS and JWST/NIRSpec redshifts, to build an improved lens model of \CL, that includes the largest sample of multiple images used to date. In addition, the new identification of several multiple image systems in the cluster core and around the external clumps enables a more accurate and precise reconstruction of the total mass distribution of the cluster. The lens model presented in this work plays a fundamental role in studying the intrinsic properties of the lensed high-redshift sources from current and future JWST observations of \CL.

The paper is organized as follows. \Sec \ref{sec:Data} describes the archival imaging and spectroscopic data sets, as well as the new JWST and MUSE observations, that are used to develop the lens model of \CL. In \Sec \ref{sec:model}, we detail the selection of the multiple images and cluster members, and the adopted methodology for the extended strong lensing modeling of the cluster. Our results are discussed in \Sec \ref{sec:results} and our main conclusions are summarized in \Sec \ref{sec:conclusions}.

Throughout this work, we adopt a flat Lambda cold dark matter ($\Lambda$CDM) cosmology with $\Omega_m = 0.3$ and $H_0= 70\,\mathrm{km\,s^{-1}\,Mpc^{-1}}$. Using this cosmology, a projected distance of $1\arcsec$ corresponds to a physical scale of 4.528 kpc at the \CL\ redshift of $z=0.3072$. All magnitudes are given in the AB system.

\section{Data} \label{sec:Data}
This section presents the datasets used to build the extended strong lensing model of \CL, focusing in particular on the new spectro-photometric data compared to those exploited by \citetalias{Bergamini2022}. In \Sec \ref{sec:imagingdata}, we describe the archival HST and Magellan imaging as well as the new JWST NIRCam imaging of this cluster field.
\Sec \ref{sec:dataSpec} summarizes the spectroscopic coverage of the cluster field and the new MUSE DDT observations. 

\subsection{Imaging data} \label{sec:imagingdata} 
We use deep, ancillary Magellan {\it g, r, i} imaging of \CL\ obtained with MegaCam on the Magellan 2 Clay Telescope on 2018 September 7-8 \citep[see][for an overview]{Treu2022}. The Magellan $i$-band imaging was used to anchor the NIRCam images to the Gaia-DR3 astrometric solution \citep[see][]{Paris2023}. All the coordinates in our new lens model are therefore also aligned to this World Coordinate System (WCS)\footnote{This corresponds to a difference of $\Delta({\rm R.A.,Dec.})_{\rm HFF-JWST}= (0.02\arcsec,-0.07\arcsec)$.}. 
The uniform Magellan multi-band photometry over the entire field can also be used to check the color consistency of a few cluster members lying outside the HST coverage. 

In addition, \CL\ is one of the cluster fields with the deepest high-resolution observations obtained with HST thanks to the Hubble Frontier Fields program\footnote{\url{https://archive.stsci.edu/prepds/frontier/}} \citep[HFF, Proposal ID: 13495,][]{Lotz_2017HFF} and other ancillary data from previous HST observational campaigns. The BUFFALO survey \citep[Beyond Ultra-deep Frontier Fields And Legacy Observations,][]{Steinhardt2020} has since provided an extended but shallower coverage of the cluster field. In this work, we make use of the HFF and BUFFALO HST mosaics that are described by \citetalias{Bergamini2022}. 

Finally, we exploit the new JWST NIRCam imaging of \CL\ obtained within the GLASS-JWST program ERS-1324 \cite[P.I.: Treu,][]{Treu2022}, the UNCOVER (Ultradeep NIRSpec and NIRCam ObserVations before the Epoch of Reionization) Cycle 1 Treasury program \cite[GO-2561, co-P.I.s: Labbé and Bezanson,][]{Bezanson2022}, and the DDT program 2756 (P.I.: Chen). The footprints from these three observational programs are shown in \Fig \ref{fig:TotClusterCM} in orange, blue, and red, respectively. The fields of view, the filters, and the data reduction are detailed in \citet{Merlin2022} and \citet{ Paris2023}. Briefly, the resulting JWST/NIRCam coverage of the cluster from the three programs includes observations with eight different filters (F090W, F115W, F150W, F200W, F277W, F356W, F410M, F444W), over an area of $\rm 46.5 ~arcmin^2$ with a $0.031\arcsec$ pixel scale. The final $5\sigma$ magnitude limit of the images ranges from $\sim28.6$ AB to $\sim30.2$ AB, depending on the location and filters. The reduced NIRCam images and the associated multi-wavelength catalogs are made publicly available\footnote{\url{https://glass.astro.ucla.edu/ers/external_data.html}}\textsuperscript{,}\footnote{\url{https://archive.stsci.edu/doi/resolve/resolve.html?doi=10.17909/kw3c-n857}}. 
The photometric redshifts are computed by means of the Spectral Energy Distribution (SED) fitting code zphot.exe \citep{Fontana2000}. Using the same method and assumptions as in \citet{Merlin2021} and \citet{Santini2023}, we have built the stellar library by adopting the \citet{Bruzual_2003} models and including nebular emission lines according to \citet{Castellano2014} and \citet{Schaerer2009}.

\subsection{Spectroscopic data} \label{sec:dataSpec}

The \CL\ cluster field counts with an extensive spectroscopic coverage obtained with several ground and space-based facilities, see \citetalias{Bergamini2022} for a detailed overview. Briefly, \CL\ was observed with the wide-field VIsible Multi-Object Spectrograph (VIMOS) as part of the ESO Large Program 169.A-0595 \citep[P.I.: Böhringer,][]{Braglia2009}, the AAOmega multi-object spectrograph on the 3.9m Anglo-Australian Telescope \citep{Owers2011}, and the HST WFC3/IR grism through
the HST GO program GLASS\footnote{\url{archive.stsci.edu/prepds/glass/}} \citep[][]{Treu_2015, Schmidt_2014}.
The $\rm \sim4 ~arcmin^2$ central region of the galaxy cluster (green footprints in \Fig \ref{fig:TotClusterCM}) was then targeted with the MUSE integral field spectrograph, mounted on the Very Large Telescope  \citep[VLT, ][]{Bacon_MUSE}, within the GTO Program 094.A-0115 (P.I.: Richard). The data consists of five MUSE pointings, each with total exposure times raging from 2 to 5 hours \citep[][\citetalias{Bergamini2022}]{Mahler_2018, Richard2021}, allowing for the spectroscopic confirmation of a large number of multiple image systems and cluster members. 
 
Besides these archival data, additional VLT/MUSE spectroscopy within the GLASS-JWST NIRCam fields was recently acquired through the ESO DDT program 109.24EZ.001 (co-P.I.s: Mason, Vanzella) on the nights of July 28 and August 20 2022 \citep[see also][for further details]{Prieto-Lyon2022}. The new data comprise 5 pointings of $\rm \sim1 ~arcmin^2$ each with a total exposure time of 1 hour per pointing (see the green footprints in \Fig \ref{fig:TotClusterCM}). 
Four of the MUSE pointings overlap with the NIRCam fields from the GLASS-JWST ERS program while the fifth targets a prominent cluster sub-structure at a distance between $\sim 600 - 775 ~ \rm kpc$ from the core of \CL\ (\citetalias{Bergamini2022}). 
The extended HST imaging from the BUFFALO program revealed several strong lensing features around two bright cluster members, subsequently confirmed by the new JWST imaging (\citetalias{Furtak2022}).

The MUSE data cubes have been reduced and analyzed following  \citet{Caminha_macs0416, Caminha_macs1206, Caminha_2019}, using the standard reduction pipeline \citep[version 2.8.5,][]{Weilbacher2020}. 
The ``autocalibration'' method and the Zurich Atmosphere Purge \citep[ZAP,][]{Soto2016} are then applied to improve the overall data reduction. The five pointings have a mean full width at half maximum (FWHM) value of $0.75\arcsec$, and the pointing covering the cluster sub-structure has a value of FWHM$=0.86\arcsec$.
 
The one-dimensional spectra of all the HST detected objects are extracted within a \textbf{$0.8\arcsec$} radius circular aperture, whereas custom apertures are considered for faint sources, based on their estimated morphology from the HST imaging (see \Sec \ref{sec:imagingdata}). For sources with no HST detection (i.e., HST-dark objects) we visually inspect the continuum subtracted data-cubes to identify emission lines. 
We exploit spectral templates, as well as the identification of emission lines, to construct the new redshift catalogs. The reliability of each redshift measurement is then quantified with the following quality flag (QF) assignments \citep[see also][]{Balestra_2016, Caminha_2019}: ``tentative'' (QF = 1), ``likely'' (QF = 2), ``secure'' (QF = 3), and ``based on a single emission line'' (QF = 9).

The full spectroscopic sample from the MUSE DDT program contains 313 reliable (i.e., QF $\ge$ 2) redshift measurements, of which 12 are stars, 12 are foreground galaxies ($z<0.28$), 82 are cluster members ($0.28\leq z\leq 0.34$) and 207 are background objects ($z>0.34$). The redshift distributions of the spectroscopic cluster members and background galaxies from the MUSE DDT observations are shown in Figures \ref{fig:hist_CM} and \ref{fig:hist_IM}, respectively. We note that 42 UV-faint $z\sim3-7$ galaxies have been published by \citet{Prieto-Lyon2022}. The full catalog is presented in \T \ref{tab:ddtcatalog}. \\

\begin{figure}[t!]
        \centering
        \includegraphics[width=1\linewidth]{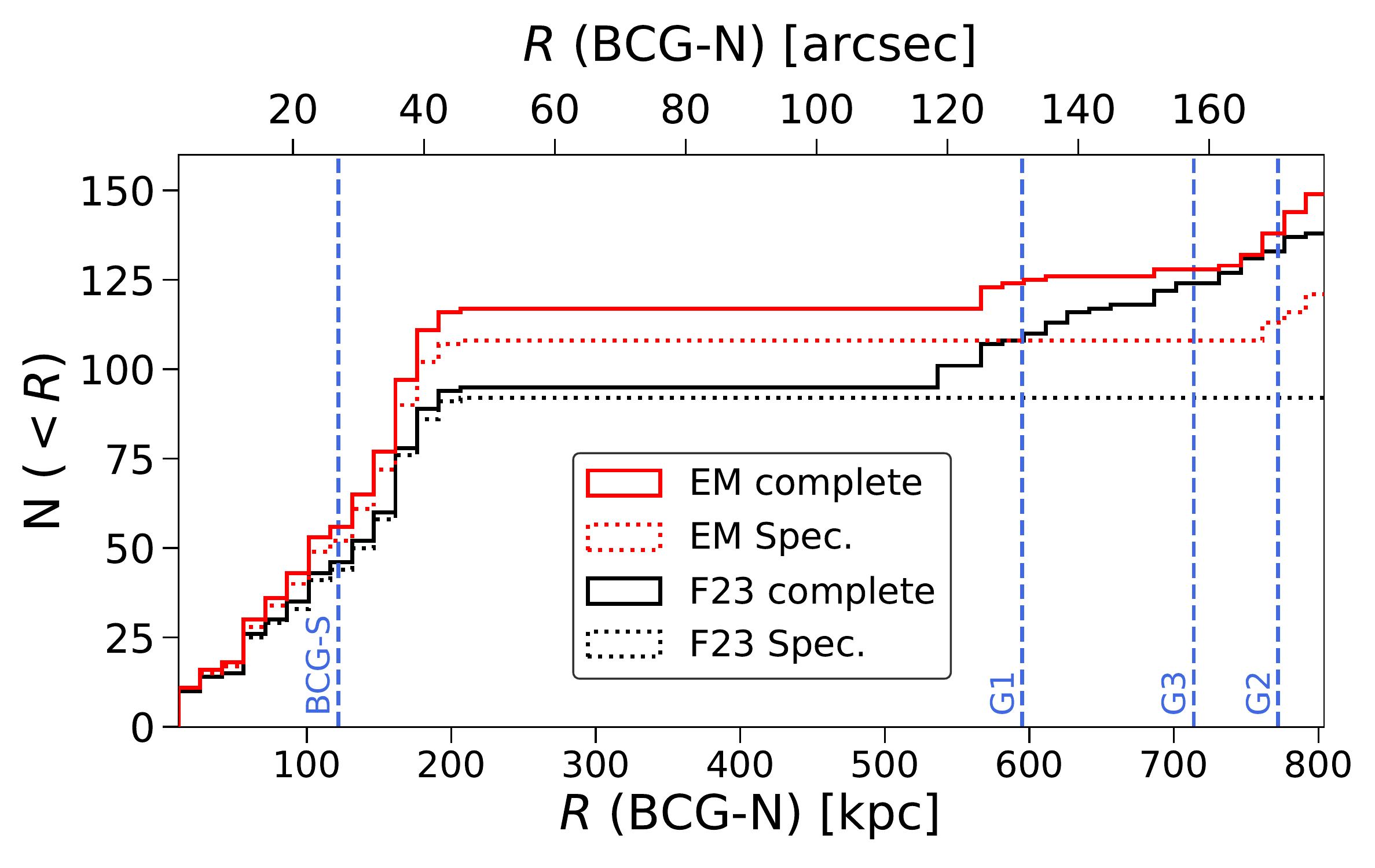}
        \caption{Cumulative distributions of the distances of the multiple images of \CL\ from the BCG-N. We show in solid red the distribution of the images used as constraints in this work (149 multiple images in total), and in dotted red the distribution of the 121 spectroscopic images. The distributions of the spectroscopic (90 multiple images, coinciding with those used by \citetalias{Bergamini2022} in addition to the two images JD1A and JD1B in \T \ref{tab:multiple_images}) and of the complete (138 multiple images) samples of multiple images used in \citetalias{Furtak2022} are shown in black. The distances of the other bright galaxies in the field (i.e., BCG-S, G1, G2, and G3) are indicated with vertical dashed blue lines.}
        \label{fig:images}
\end{figure}

\begin{figure*}[h!]
        \centering
        \includegraphics[width=1\linewidth]{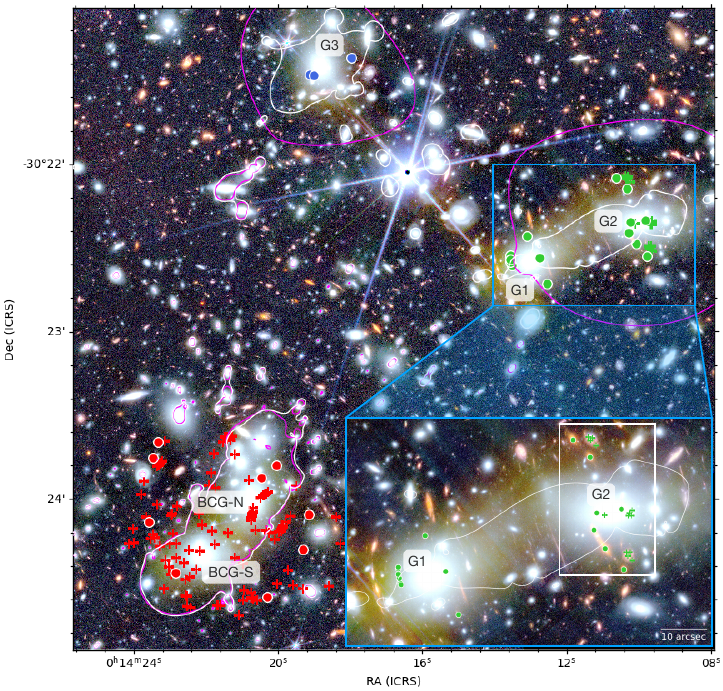}
        \caption{JWST color image (R: F444W+F356W, G: F277W+F200W, B: F150W+F115W) of the galaxy cluster \CL\ showing the positions of the 149 multiple images, from 50 background sources, used to constrain the lens model. The colored crosses show the positions of the 121 spectroscopically confirmed multiple images, while the colored circles are the 28 non-spectroscopic multiple images used in the lens model. Red, green, and blue colors highlight the images concentrated around the cluster BCGs (117 multiple images from 39 background sources), the two bright galaxies G1 and G2 (29 multiple images from 10 background sources), and the galaxy G3 (3 multiple images from a single background source), respectively. The critical lines at $z_s=6$ obtained from the best-fit lens model are drawn in white, while those from the previous model by \citetalias{Bergamini2022} are shown in magenta as a comparison.}
        \label{fig:TotClusterImage}
\end{figure*}

\begin{figure*}[t!]
        \centering
        \includegraphics[width=1\linewidth]{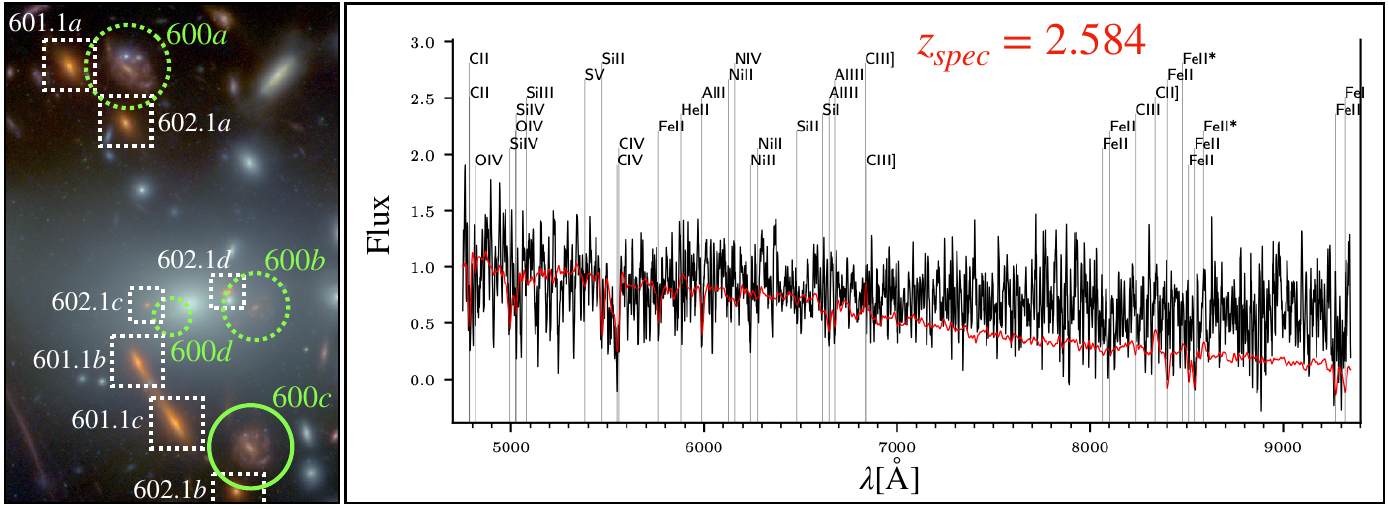}
        \caption{MUSE spectrum of the multiple image 600c (containing the multiply imaged substructures 600.1c, 600.3c, 600.4c, and 600.5c) overlaid to a star-forming template from \citet{Talia2012} (red line). This system is a face-on spiral at $z=2.584$ marked as a solid green circle in the left panel (corresponding to the white box in \Fig \ref{fig:TotClusterImage} around G2). Three other images of the system (600a,b,d) are indicated as dashed green circles; white dotted squares show the multiple images 601.1a,b,c and 602.1a,b,c,d with no spectroscopic redshifts (see \Fig \ref{fig:zlens} and \T \ref{tab:multiple_images}). }
        \label{fig:spec}
\end{figure*}

\section{Extended strong lensing model} \label{sec:model}
In this section we describe the new strong lensing model of \CL\ making use of the new imaging and spectroscopic datasets presented in \Sec \ref{sec:Data}.
In particular, thanks to the deep JWST imaging and the MUSE DDT follow-up observations, we are able to expand the samples of multiple images and spectroscopic cluster members compared to the model presented by \citetalias{Bergamini2022}.
We use the publicly available strong lensing modeling pipeline \LT \footnote{\url{https://projets.lam.fr/projects/lenstool/wiki}} \citep[][]{Kneib_lenstool, Jullo_Kneib_lenstool} that reconstructs the total mass distribution of a galaxy cluster implementing a Bayesian technique. We note that the lens model adopts the new JWST astrometric grid, tied to Gaia DR3 (see \Sec \ref{sec:imagingdata}).

\subsection{Parameterization of the cluster total mass} \label{sec:MassParametrization}
Thanks to the identification of a large number of multiple images in the external clumps (see \Sec \ref{sec:multiple_images}), we are able to improve the characterization of the total mass distribution of \CL\ compared to our previous, HST-based strong lensing model by \citetalias{Bergamini2022}. The total mass parameterization of the cluster core is the same as the one adopted by \citetalias{Bergamini2022}. We thus refer the reader to that publication for a detailed description and provide hereafter a brief summary.
Within the parametric \LT\ software \citep{Jullo_lenstool}, the total mass of the lens is separated into several components, where the following mass contributions are considered in this analysis:

\begin{equation}
    \label{eq.: pot_dec}
    \phi_{tot}= \sum_{i=1}^{N_h}\phi_i^{halo}+\sum_{j=1}^{N_{BCG}}\phi_j^{BCG}+\sum_{k=1}^{N_g}\phi_k^{gal}+\sum_{l=1}^{N_s}\phi_l^{ENV}.
\end{equation}
 
The cluster and subhalo mass components ($\phi^{halo}$, $\phi^{BCG}$ and $\phi^{gal}$) are parameterized using dual pseudo-isothermal elliptical mass distributions \citep[dPIEs,][]{Limousin_lenstool, Eliasdottir_lenstool, Bergamini_2019}. The dPIE profile is defined by seven free parameters: the position ($x$, $y$); the ellipticity (defined as $e=\frac{a^2-b^2}{a^2+b^2}$, where $a$ and $b$ are the values of the major and minor semiaxes, respectively); the position angle $\theta$, computed counterclockwise from the west direction; the central velocity dispersion $\sigma_{0}$\footnote{\LT\ adopts a scaled version of this quantity, identified as $\sigma_{LT}$, such that $\sigma_{LT}=\sigma_{0} \sqrt{2/3}$}; the core radius $r_{core}$; and the truncation radius $r_{cut}$. 

The large-scale dark matter component, $\phi^{halo}$, consists of two non-truncated elliptical dPIEs, which are initially centered on the two Brightest Cluster Galaxies (BCGs; labeled BCG-N and BCG-S) but are free to move within a small range around their positions (corresponding to square regions of sizes of about $ \rm 45\,kpc$ and $\rm 90\,kpc$ for the BCG-N and BCG-S, respectively; see \T \ref{table:inout_lensing}).
As pointed out by \citetalias{Bergamini2022}, the observed positions of several multiple image systems with small angular separation from the BCG-N and BCG-S are better reproduced when the parameters describing their mass contribution -- $\phi^{BCG}$  and ellipticity -- are optimized outside the scaling relations adopted for the other cluster members, as discussed below. BCGs are known to be uncommon galaxies and their independent modeling highlights the importance of the two BCGs for an accurate lensing analysis.

Of the 177 cluster member galaxies (see \Sec \ref{sec:cluster_members}), 172 are modeled using circular dPIEs with a vanishing core radius and scaled with total mass-to-light ratios increasing with their HST F160W luminosities, i.e., a relation that is compatible with the so-called tilt of the Fundamental Plane \cite[e.g.,][]{Faber1987, Bender1992}, as done by \citetalias{Bergamini2022} (see their \Eq 4).
The two free parameters in the lens model are then the values of the velocity dispersion and truncation radius of a reference galaxy, corresponding to the BCG-N. We report in \T \ref{table:inout_lensing} the values of the relevant parameters of the scaling relations. \citetalias{Bergamini2022} further exploited the MUSE datacube to measure the line-of-sight stellar velocity dispersion for 85 member galaxies, down to $m_{\rm F160W} \sim 22$, allowing for an independent calibration of the slopes of the velocity dispersion and truncation radius scaling relations (see \citetalias{Bergamini2022} for details).

As previously mentioned, the main difference with respect to the strong lensing analysis presented by \citetalias{Bergamini2022} is the modeling of the cluster infalling regions, $\phi^{ENV}$. Due to the lack of secure multiple image systems around the massive structures in the northwestern region of the cluster prior to the JWST/NIRCam imaging, the \citetalias{Bergamini2022} reference \texttt{LM-model} included a simple description of the mass contribution from three external clumps, associated with the three brightest galaxies (G1, G2, and G3), as singular isothermal sphere (SIS) profiles.
The identification of 29 (3) multiple images around the galaxies G1-G2 (G3) allows now for a more accurate total mass reconstruction of these infalling structures, $\phi^{ENV}$, which is decomposed into the following mass contributions:

\begin{equation}
    \label{eq.: pot_dec}
    \phi^{ENV}= \phi_{G1-G2}^{halo} + \phi_{G3}^{halo} + \phi_{G1} + \phi_{G2} + \phi_{G3}
\end{equation}

The first two terms correspond to the cluster component that is modeled as two cluster-scale non-truncated elliptical dPIEs: one associated with the galaxies G1 and G2 and allowed to move within a large square region containing the two galaxies (see \T \ref{table:inout_lensing}, corresponding to a region of about $ \rm 270\,kpc\times180\,kpc$), and a second one centered on the position of the galaxy G3. 
The parameters describing the mass contribution (i.e., the velocity dispersion and cut radius) of the three galaxies G1, G2 and G3 are individually optimized, within large flat priors. In addition, the ellipticity parameters of G1 and G2 are left free to vary (see \T \ref{table:inout_lensing}).
As mentioned in \Sec \ref{sec:cluster_members}, the parametric description of the cluster outskirsts is further enhanced by the inclusion of a sample of spectroscopic cluster members from the new MUSE/DDT observations (see Figure \ref{fig:TotClusterCM} and \Sec \ref{sec:cluster_members}).

Finally, the redshifts of the 8 photometric multiple image systems are also optimized in the lens model within uninformative flat priors. We note that the redshift value of all the images of Sys-200 (i.e., A200.1a,b, B200.2a,b, and C200.3a,b) is imposed to be equal in the optimization of the lens model (and therefore corresponding to a single free parameter).

The priors assumed for the values of the parameters of the mass profiles included in our reference lens model are reported in the upper part of \T \ref{table:inout_lensing}, while the optimized values are provided in the bottom panel. 
The best-fitting values of the model parameters that describe the total mass distribution of the lens are obtained by minimizing on the image plane the distance between the observed and the model-predicted point-like positions of the multiple images through a $\chi^2$ function (see \Eq 1 in \citetalias{Bergamini2022}).
Following \citetalias{Bergamini2022}, we assign an initial positional uncertainty to each image depending on the value of the \textit{Positional Quality Flag} (QP), that are given in \T \ref{tab:multiple_images}.
These values are then rescaled, prior to the sampling of the posterior distributions, in order to obtain a $\chi^2$ value close to the number of degrees of freedom (dof) in the model, defined as: $\mathrm{dof}=2\times [N^{tot}_{im} - N_{fam}] - N_{freepar}=N_{con} - N_{freepar}$, where $N^{tot}_{im}$ and $N_{fam}$ refer, respectively, to the total number of multiple images and families included in the lens model. $N_{con}$ and $N_{freepar}$ are the number of model constraints and free parameters, respectively.
The total mass model of \CL\ has $N_{freepar}=50$ free parameters, including the 8 optimized redshifts of the non-spectroscopic systems, leading to 148 dof. This model represents an extension of that presented by \citetalias{Bergamini2022}, and thus is labeled as such in the following (\texttt{Extended Model} or EM).

\begin{figure}[t!]
        \centering
        \includegraphics[width=1\linewidth]{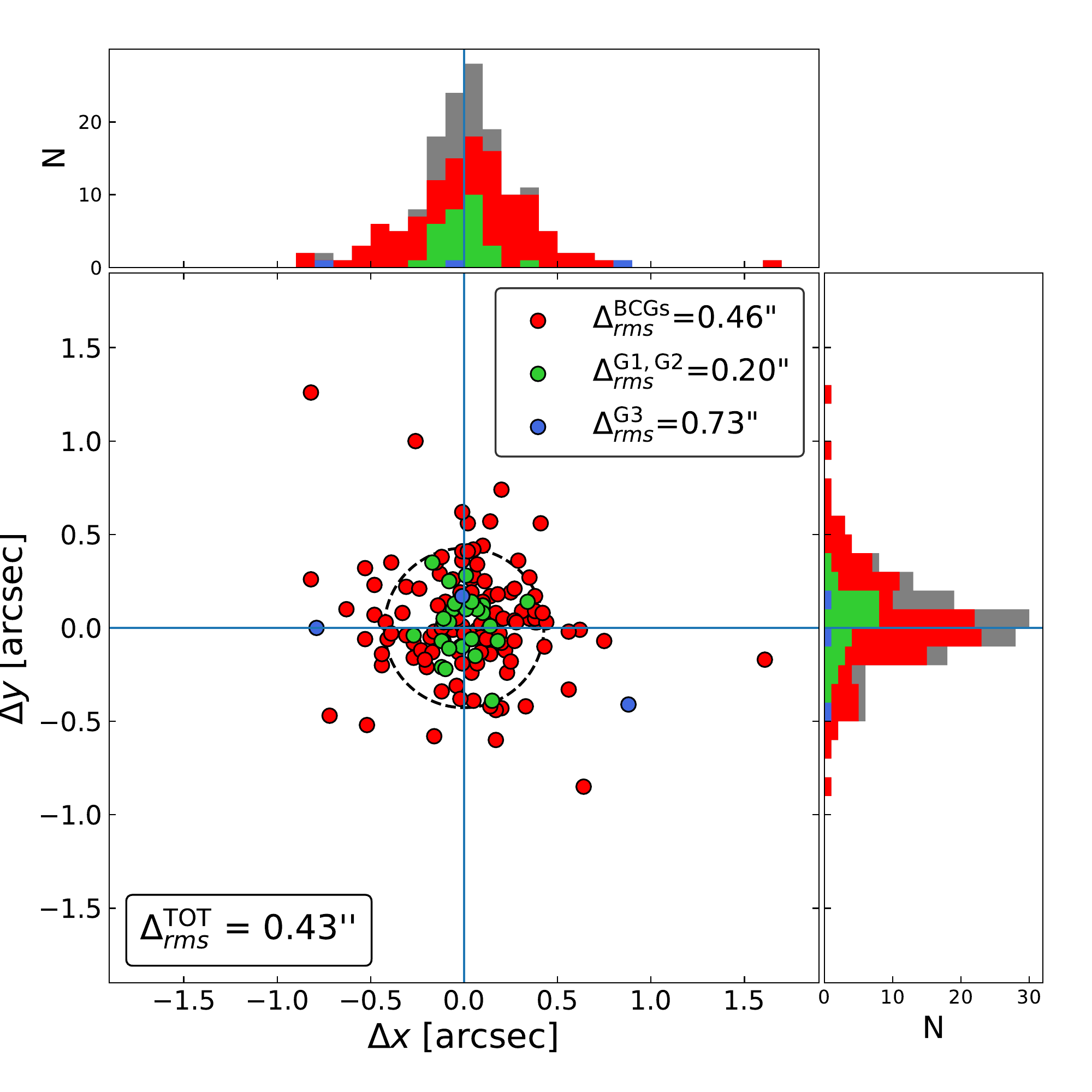}
        \caption{Displacements, $\boldsymbol{\Delta}_i$, along the $x$ and $y$ directions between the observed and model predicted positions of the 149 multiple images (from 50 background sources) included in the lens model. The lens model is characterized by a $\Delta^{\rm TOT}_{rms}=0.43\arcsec$ corresponding to the radius of the black dashed circle. The 117 multiple images, from 39 background sources, concentrated around the two BCGs are plotted in red and have a $\Delta_{rms}=0.46\arcsec$. Similarly, the 29 multiple images (from 10 background sources) forming around the two bright galaxies G1 and G2 (with $\Delta_{rms}=0.20\arcsec$), and the 3 multiple images (from a single background source) around the galaxy G3 (with $\Delta_{rms}=0.73\arcsec$) are plotted in green and blue, respectively. The gray histograms refer to the complete sample of multiple images.}
        \label{fig:RMS}
\end{figure}

\begin{table*}[ht!]   
    \tiny
	\def\arraystretch{2.3}
	\centering          
  \caption{Input and output parameters of the extended lens model for the galaxy cluster \CL\ presented in this work.}
	\begin{tabular}{|c|c|c|c|c|c|c|c|c|}
	    \cline{3-9}
		\multicolumn{2}{c|}{} & \multicolumn{7}{c|}{ \textbf{Input parameter values and assumed priors}} \\
		\cline{3-9}
		  \multicolumn{2}{c|}{} & \boldmath{$x\, \mathrm{[arcsec]}$} & \boldmath{$y\, \mathrm{[arcsec]}$} & \boldmath{$e$} & \boldmath{$\theta\ [^{\circ}]$} & \boldmath{$\sigma_{LT}\, \mathrm{[km\ s^{-1}]}$} & \boldmath{$r_{core}\, \mathrm{[arcsec]}$} & \boldmath{$r_{cut}\, \mathrm{[arcsec]}$} \\ 
          \hline

		  \multirow{4}{*}{\rotatebox[origin=c]{90}{\textbf{Cluster-scale halos}}} 
		  
		  & \boldmath{$1^{st}$} \bf{Cluster Halo} & $-5.0\,\div\,5.0$ & $-5.0\,\div\,5.0$ & $0.0\,\div\,0.9$ & $0.0\,\div\,180.0$ & $300\,\div\,1500$ & $0.0\,\div\,30.0$ & $2000.0$ \\
		  
		  & \boldmath{$2^{nd}$} \bf{Cluster Halo} & $-27.9\,\div\,-7.9$ & $-30.0\,\div\,-10.0$ & $0.0\,\div\,0.9$ & $0.0\,\div\,90.0$ & $300\,\div\,1500$ & $0.0\,\div\,30.0$ & $2000.0$  \\

		  \cline{2-9}

		  & \bf{G1,G2 Halo} & $90.0\,\div\,150.0$ & $80.0\,\div\,120.0$ & $0.0\,\div\,0.9$ & $0.0\,\div\,180.0$ & $300\,\div\,1500$ & $0.0\,\div\,30.0$ & $2000.0$ \\
		  
		  & \bf{G3 Halo} & $24.2$ & $155.8$ & $0.0\,\div\,0.9$ & $0.0\,\div\,180.0$ & $100\,\div\,1500$ & $0.0\,\div\,30.0$ & $2000.0$ \\

          \hline
          \multicolumn{1}{c}{}
          \\[-5ex]

          \hline

		  \multirow{6}{*}{\rotatebox[origin=c]{90}{\textbf{Subhalos}}}

		  & \bf{BCG-N} & $0.0$ & $0.0$ & $0.0\,\div\,0.9$ & $0.0\,\div\,180.0$ & $200\,\div\,400$ & $0.0001$ & $0.1\,\div\,50.0$ \\
		  
  		& \bf{BCG-S} & $-17.9$ & $-20.0$ & $0.0\,\div\,0.9$ & $0.0\,\div\,180.0$ & $200\,\div\,400$ & $0.0001$ & $0.1\,\div\,50.0$ \\

            & \bf{G1} & $99.4$ & $85.9$ & $0.0\,\div\,0.9$ & $0.0\,\div\,180.0$ & $150\,\div\,350$ & $0.0001$ & $0.1\,\div\,50.0$ \\

            & \bf{G2} & $138.3$ & $99.8$ & $0.0\,\div\,0.9$ & $0.0\,\div\,180.0$ & $150\,\div\,350$ & $0.0001$ & $0.1\,\div\,50.0$ \\

            & \bf{G3} & $24.2$ & $155.8$ & $0.0$ & $0.0$ & $150\,\div\,400$ & $0.0001$ & $0.1\,\div\,50.0$ \\
		  
		  \cline{2-9}
		  
		  & \bf{Scaling relations} & $\boldsymbol{N_{gal}=}172$
		  & $\boldsymbol{m_{\mathrm{F160W}}^{ref}=}17.34$
		  & $\boldsymbol{\alpha=}0.40$ & $\boldsymbol{\sigma_{LT}^{ref}=}190\,\div\,300$ & $\boldsymbol{\beta_{cut}=}0.41$ & $\boldsymbol{r_{cut}^{ref}=}0.5\,\div\,10.0$ & $\boldsymbol{\gamma=}0.20$\\
		  
		  \hline
		 
	\end{tabular}
	\\[6ex]
	\begin{tabular}{|c|c|c|c|c|c|c|c|c|}
	    \cline{3-9}
		\multicolumn{2}{c|}{} & \multicolumn{7}{c|}{ \textbf{Optimized output parameters}} \\
		\cline{3-9}
		  \multicolumn{2}{c|}{} & \boldmath{$x\, \mathrm{[arcsec]}$} & \boldmath{$y\, \mathrm{[arcsec]}$} & \boldmath{$e$} & \boldmath{$\theta\ [^{\circ}]$} & \boldmath{$\sigma_{LT}\, \mathrm{[km\ s^{-1}]}$} & \boldmath{$r_{core}\, \mathrm{[arcsec]}$} & \boldmath{$r_{cut}\, \mathrm{[arcsec]}$} \\ 
          \hline

		  \multirow{4}{*}{\rotatebox[origin=c]{90}{\textbf{Cluster-scale halos}}} 
		  
		  & \boldmath{$1^{st}$} \bf{Cluster Halo} & $-1.8^{+0.6}_{-0.6}$ & $-3.5^{+1.4}_{-1.1}$ & $0.4^{+0.1}_{-0.1}$ & $79.0^{+5.3}_{-7.0}$ & $558^{+33}_{-35}$ & $9.3^{+1.1}_{-1.1}$ & $2000.0$ \\
		  
		  & \boldmath{$2^{nd}$} \bf{Cluster Halo} & $-18.7^{+0.5}_{-0.4}$ & $-16.3^{+0.5}_{-0.5}$ & $0.4^{+0.1}_{-0.1}$ & $55.2^{+2.8}_{-2.7}$ & $603^{+25}_{-25}$ & $8.1^{+0.6}_{-0.6}$ & $2000.0$  \\
		  
		  \cline{2-9}

    	& \bf{G1,G2 Halo} & $130.2^{+2.2}_{-2.8}$ & $97.9^{+0.8}_{-1.1}$ & $0.7^{+0.1}_{-0.1}$ & $23.8^{+2.1}_{-1.9}$ & $781^{+35}_{-38}$ & $16.8^{+2.0}_{-2.0}$ & $2000.0$  \\

           & \bf{G3 Halo} & $24.2$ & $155.8$ & $0.1^{+0.2}_{-0.1}$ & $14.1^{+25.7}_{-67.1}$ & $799^{+40}_{-37}$ & $22.9^{+3.8}_{-4.4}$ & $2000.0$  \\

          \hline
          \multicolumn{1}{c}{}
          \\[-5ex]

          \hline

		  \multirow{6}{*}{\rotatebox[origin=c]{90}{\textbf{Subhalos}}}

		  & \bf{BCG-N} & $0.0$ & $0.0$ & $0.6^{+0.1}_{-0.2}$ & $87.5^{+5.8}_{-8.1}$ & $266^{+15}_{-12}$ & $0.0001$ & $31.8^{+11.6}_{-12.1}$ \\
		  
  		& \bf{BCG-S} & $-17.9$ & $-20.0$ & $0.7^{+0.1}_{-0.1}$ & $23.7^{+3.3}_{-3.3}$ & $296^{+12}_{-14}$ & $0.0001$ & $44.9^{+3.7}_{-7.1}$ \\

            & \bf{G1} & $99.4$ & $85.9$ & $0.3^{+0.3}_{-0.2}$ & $17.3^{+21.5}_{-54.6}$ & $232^{+16}_{-18}$ & $0.0001$ & $29.0^{+11.0}_{-9.7}$ \\

            & \bf{G2} & $138.3$ & $99.8$ & $0.4^{+0.2}_{-0.2}$ & $-14.0^{+14.7}_{-19.8}$ & $231^{+15}_{-16}$ & $0.0001$ & $23.3^{+13.2}_{-11.5}$ \\

            & \bf{G3} & $24.2$ & $155.8$ & $0.0$ & $0.0$ & $332^{+38}_{-38}$ & $0.0001$ & $13.4^{+8.4}_{-6.3}$ \\
		  
		  \cline{2-9}
		  
		  & \bf{Scaling relations} & $\boldsymbol{N_{gal}=}172$
		  & $\boldsymbol{m_{\mathrm{F160W}}^{ref}=}17.34$
		  & $\boldsymbol{\alpha=}0.40$ & $\boldsymbol{\sigma_{LT}^{ref}=}246^{+9}_{-10}$ & $\boldsymbol{\beta_{cut}=}0.41$ & $\boldsymbol{r_{cut}^{ref}=}9.3^{+0.5}_{-1.0}$ & $\boldsymbol{\gamma=}0.20$\\
		  
		  \hline
		 
	\end{tabular}
	\\[6ex]

    \tablecomments{The $\boldsymbol{x}$ and $\boldsymbol{y}$ coordinates are expressed in arcseconds with respect to the position of the BCG-N, at R.A.=3.586244, Dec.=$-$30.400151. In the {\it input parameter table}, single numbers are quoted for parameters that are kept fixed during the model optimization. When two values (separated by the $\div$ symbol) are quoted for a parameter, they correspond to the boundaries of the flat prior assumed in the model. $\boldsymbol{N_{gal}}$ is the number of cluster member galaxies optimized through the scaling relations (see \Eq 4 in \citetalias{Bergamini2022}). In the last line of each table, we report the reference magnitude, $\boldsymbol{m_{\mathrm{F160W}}^{ref}}$, and the remaining parameters of the scaling relations. 
    In the {\it output parameter table}, we quote the optimized values of the model parameters. For each free parameter, we quote the median and the 16-th, and 84-th percentiles from the model marginalized posterior distributions. The posterior distributions of the angular parameter $\mathbf{\theta}$ of the G3 halo and of the galaxies G1 and G3 are re-mapped to an angular interval between $-$90 and 90 degrees before computing the percentiles.
    }    

	\label{table:inout_lensing}

\end{table*}

\subsection{Multiple image catalog} \label{sec:multiple_images}
This work extends the multiple image catalog presented by \citetalias{Bergamini2022}, which was based on the deep HST imaging and MUSE observations of the cluster core. The \citetalias{Bergamini2022} sample consisted of 90 multiple images from 30 background sources, spanning a redshift range between $z = 1.69$ and $z = 5.73$. For further details, we refer the reader to \T A.1 of \citetalias{Bergamini2022}. 
By exploiting the ancillary and the recently obtained JWST/NIRCam imaging and new VLT/MUSE data, we identify additional multiple images both in the cluster core and in the external clumps.

Our new strong lensing model includes 149 multiple images from 50 background sources, of which 121 are spectroscopically confirmed and span an extended redshift range between $z = 1.03$ and $z = 9.76$ (see \Fig \ref{fig:hist_IM}). This represents an increase of $\sim66\%$ compared to the \citetalias{Bergamini2022} sample. 
The cumulative distribution of the distances of both the spectroscopic and photometric multiple images included in the lens model from the BCG-N is shown in \Fig \ref{fig:images} (red solid line), and compared to the one used by \citetalias{Furtak2022} (black solid line). The complete sample presented in this work includes 11 more multiple images than that from \citetalias{Furtak2022}, and thus represents the largest set of constraints included in a lens model of \CL\ to date. The set of spectroscopic images is especially noteworthy. In more general terms, \CL\ is the second cluster with the largest sample of secure multiple images after the lens cluster MACS J0416.1$-$2403, that currently counts 237 \citep{Bergamini2023b}.
For consistency, we present in \T \ref{tab:multiple_images} the properties (in the JWST-based astrometry) of the complete sample of multiple images used as constraints in this work, included those presented by \citetalias{Bergamini2022}. 
The multiple image positions are shown in \Fig \ref{fig:TotClusterImage}, where the crosses (circles) denote the spectroscopic (photometric) images.
We briefly describe below the identification of the new multiple images with respect to those from \citetalias{Bergamini2022}.

- \textit{Cluster core:}
Following \citetalias{Bergamini2022}, we re-analyze the MUSE datacube, extracting the spectra of sources predicted by the lens model and measuring the redshift of several multiple images using faint emission lines and the cross-correlation with different templates. 
We thus include 14 additional spectroscopically confirmed multiple images in the cluster core, from 5 systems. These new images are flagged by a star symbol in \T \ref{tab:multiple_images}. 

Recent JWST observations have allowed for the spectroscopic confirmation of two additional multiple images in the cluster core as part of the GLASS-JWST ERS program (identified by an asterisk in \T \ref{tab:multiple_images}).
While image 3c is too faint for a secure redshift measurement with VLT/MUSE, JWST/NIRISS spectroscopy enables the detection of [OII]$\lambda$3727,3729 at the expected wavelength position of 1.856 $\rm \mu m$, and consistent with the counterimages 3a and 3b \citep{Vanzella2022, Lin2023}.
The $z_{phot}=9.8$ triply lensed candidate system reported by \citet{Zitrin2014} was observed with NIRSpec prism spectroscopy (DDT 2756; P.I.: Chen) resulting in a spectroscopic redshift measurement of $9.756^{+0.017}_{-0.007}$ of the image JD1B \citep{Roberts-Borsani2022}. The other two images, JD1A and JD1C, which are not yet spectroscopically confirmed, are considered at the same redshift as JD1B, consistent with their photometric redshifts. 

The list of multiple images is further enhanced in the cluster core by including 3 photometric strongly lensed sources, with a total of 9 multiple images, securely identified in the JWST/NIRISS and JWST/NIRCam imaging from the GLASS-JWST-ERS program. 
In particular, Sys-53 consists of a triply imaged active galactic nucleus (AGN) candidate analyzed in~\cite{Furtak2022_2}. The very peculiar color and compactness of the source make the association of the three multiple images of the system particularly secure. 

The final sample in the main cluster core includes 117 multiple images from 39 background sources, of which 108 are spectroscopically confirmed. 

- \textit{External clumps:}
The main improvement with respect to \citetalias{Bergamini2022} is the identification of a large number of multiple images around the galaxies G1 and G2 in the northwestern external clump. While the ancillary shallow HST imaging of the cluster outskirts hinted at the presence of strong lensing features, the recent deep, high-resolution JWST/NIRCam data clinched the identification of several photometric systems. 
We identify 29 multiple images from 10 background sources around the galaxies G1 and G2, and 3 multiple images from a single background source around the galaxy G3. We also include multiply lensed clumps within resolved extended sources which have shown to be particularly efficient at constraining the position of the critical curves \citep[e.g.][\citetalias{Bergamini2022}]{Grillo2016}.  
In particular, thanks to the MUSE DDT observations we are able to measure the redshift of the lensed face-on spiral galaxy (ID 600) around G2. Due to the light contamination by a star and by G2 of the images 600a and 600b, respectively, a secure spectroscopic confirmation is not currently possible for those images. The MUSE spectrum of image 600c, extracted within a circular aperture of $0.8\arcsec$ diameter, is shown in \Fig \ref{fig:spec}, yielding a redshift measurement of $2.584$. This secure redshift is based on the cross-correlation of the MUSE spectrum with several spectral templates of star-forming galaxies with appropriate rest frame UV coverage, which consistently shows a significant peak at the quoted redshift with a variation of $\Delta z=0.003$. This is, to date, the only multiple image system spectroscopically confirmed in the northwestern external clump. 

The observed image positions of the 149 multiple images are used as constraints in the new lens model, providing in total $N_{con}=198$ constraints.

\subsection{Selection of cluster galaxies} \label{sec:cluster_members}
For the selection of the cluster galaxies used in the lens model to describe the subhalo mass component, we follow the procedure of \citetalias{Bergamini2022} taking advantage of the enhanced spectroscopic coverage in the northwestern clump. Initially, cluster members are selected as the galaxies with a spectroscopic redshift in the range $0.28 \le z \le 0.34$, corresponding approximately to an interval of $\pm 6000$ km/s rest-frame velocities around the median cluster redshift $z_{CL}=0.3072$, and magnitudes brighter than $m_\mathrm{F160W}=24$ (a dedicated work on the properties of cluster members is presented by \citealt{Vulcani2023}). The larger range in redshift is motivated by the complexity of the total mass distribution of \CL, a multi-component merger \citep[see e.g.,][]{Merten2011, Owers2011}. This is clearly shown in \Fig \ref{fig:hist_CM}, where the velocity difference between the BCG-N and BCG-S is of about 4000 km/s.
The new MUSE DDT data add 82 members to the full spectroscopic sample (see \Sec \ref{sec:dataSpec}) which now includes 669 cluster galaxies, mostly based on the MUSE observations \citep[][\citetalias{Bergamini2022}]{Richard2021}. When compiling this catalog, we only consider secure or probable/likely redshift measurements from different sources (see \Sec \ref{sec:dataSpec}). The redshift distribution of the cluster members over the cluster area is shown in \Fig \ref{fig:hist_CM}, while their spatial distribution is shown in the right panel of \Fig \ref{fig:TotClusterCM}. 
The spectroscopic sample is then completed down to $m_\mathrm{F160W}=24$ by adding the photometric members identified with the convolution neural network (CNN) technique described by \citet{Angora_2020}.  
Performance tests generally show a cluster member identification rate (completeness) of $\sim\! 90\%$ with a purity of $\sim\! 95\%$, superior to those from traditional color selection methods \citep{Angora_2020}. 

The large number of cluster galaxies makes the extended lens model presented here computationally very expensive, owing to the calculation of the deflection angle induced by each individual cluster member.
With the aim of limiting the computational time increase, we reduce the number of faint cluster galaxies in the lens model and study how the positional rms value, $\Delta_{rms}$, varies when considering different magnitude threshold values for the cluster members in the F160W band. We find no significant variation in $\Delta_{rms}$ when only including in the model the galaxies with magnitudes $m_{\rm F160W}<21$ and thus, we limit the number of cluster galaxies to those brighter than $m_{\rm F160W}=21$. A similar finding is presented by \citet{Raney2021}, which shows that considering different magnitude limits for the selection of member galaxies in the lens models of two HFF clusters, between 21 and 26 in the F814W band, results in little variations in the values of $\Delta_{rms}$ or other metrics (see their \Fig 4).

With the new magnitude cut, $m_{\rm F160W}=21$, 163 spectroscopic members are included in the lens model (see the red histogram  in \Fig \ref{fig:hist_CM}), in addition to 14 non-spectroscopic galaxies, over an area of $\sim\! 45$ arcmin$^2$. The new spectroscopic cluster members are concentrated in the NW mass clumps, around G1, G2, and G3 (see \Fig \ref{fig:TotClusterCM}). As a comparison, the \citetalias{Bergamini2022} strong lensing model included 225 members with a limit 3 magnitudes fainter.

\begin{figure*}[t!]
        \centering
        \includegraphics[width=1\linewidth]{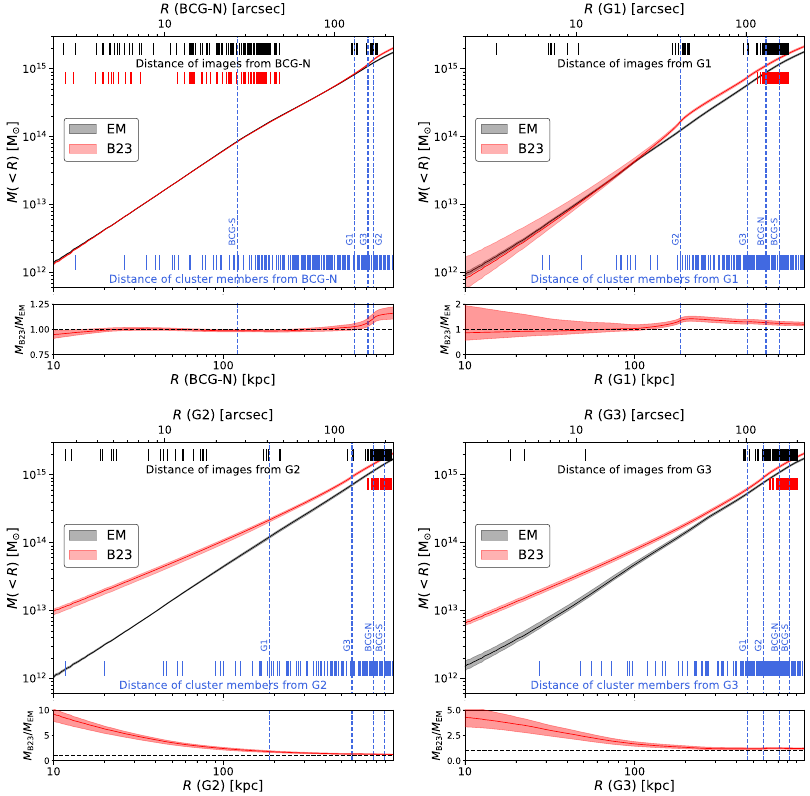}
        \caption{Cumulative projected total mass profiles of the galaxy cluster \CL\ as a function of the projected distance from the bright galaxies BCG-N (top left), G1 (top right), G2 (bottom left), and G3 (bottom right). The  mass profiles obtained from the extended lens model (EM) are drawn in black. The small vertical black and blue bars show the projected distances of the observed multiple images and of the cluster member galaxies considered in the model, respectively. We also highlight, using vertical dashed lines, the positions of the bright cluster galaxies BCG-N, BCG-S, G1, G2, and G3.
        In red, we plot the cumulative projected total mass profiles from the previous lens model by \citetalias{Bergamini2022}. The projected distances of the multiple images constraining that model are marked with small vertical red bars. At the bottom of each panel, we show the ratio between the total mass profiles obtained from the \citetalias{Bergamini2022} model and the extended model presented here.}
        \label{fig:mass}
\end{figure*}

\begin{figure*}
        \centering
        \includegraphics[width=1\linewidth]{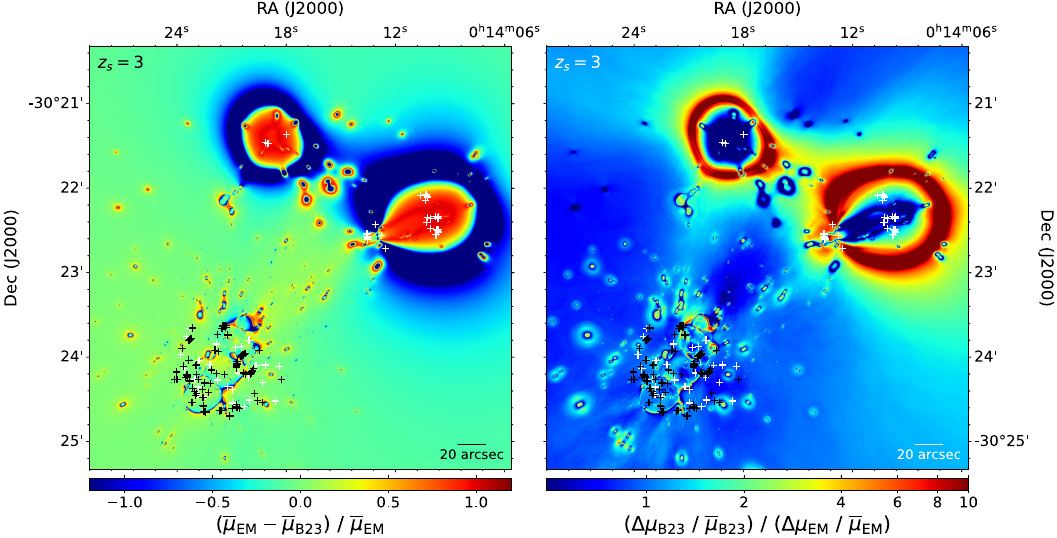}
        \caption{Comparison between the magnification factors, and related errors, obtained from the \CL\ extended lens model presented in this paper and the previous model by \citetalias{Bergamini2022}. The black crosses indicate the observed positions of the 90 multiple images used in \citetalias{Bergamini2022}. The white crosses indicate instead those for the 59 new multiple images from this work. \textit{Left:} Relative difference between the median magnification maps (see \Sec \ref{sec:results}) obtained from the extended and \citetalias{Bergamini2022} lens models. \textit{Right:} Ratio between the relative errors on the magnification maps values estimated from 500 realizations of the lens models obtained by randomly extracting parameter samples from the model MCMC chains.}
        \label{fig:magnificatio_maps}
\end{figure*}

\begin{figure*}[t!]
        \centering
        \includegraphics[width=1\linewidth]{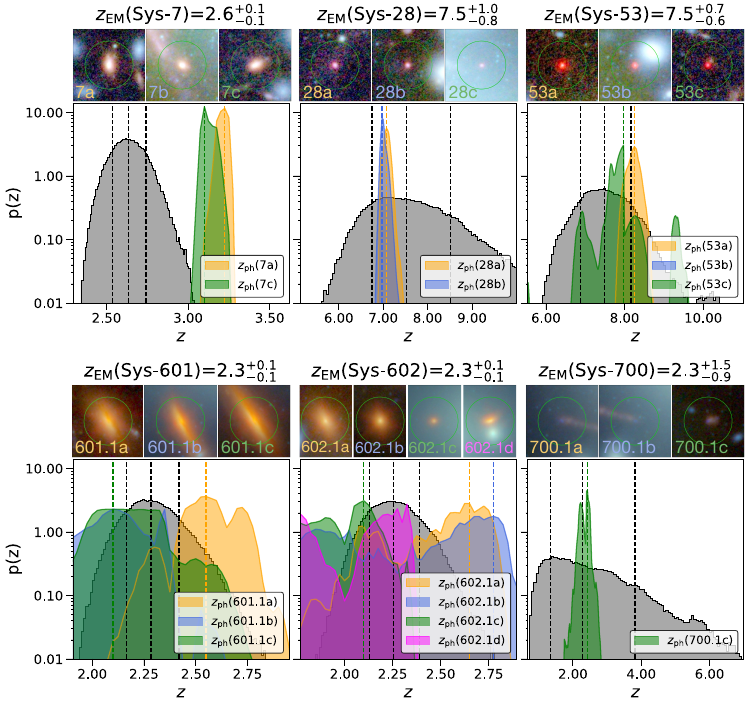}
        \caption{Probability density distributions for the redshift values of the six non-spectroscopic systems of multiple images included in the lens model. The gray histograms correspond to the marginalized probability distributions obtained from the lens model optimization. The orange, blue, green, and magenta distributions represent the photometric redshift probability density distributions measured for the multiple images of the same system. On the top of the histograms, we show color cutouts centered on the corresponding multiple images; the green circles have a radius of $1\arcsec$. The remaining two non-spectroscopic systems of multiple images considered in the lens model have optimized redshift values of $z_{\rm EM}(\mathrm{Sys}\mbox{-}200)=5.9^{+2.4}_{-2.3}$ and $z_{\rm EM}(\mathrm{Sys}\mbox{-}301)=4.1^{+1.2}_{-1.2}$. None of their images have a measured photometric redshift.}
        \label{fig:zlens}
\end{figure*}

\null\vspace{1cm}
\section{Results} \label{sec:results}
The best-fit \texttt{Extended Model} is characterized by a precision of $\boldsymbol{\Delta}_{rms}^\mathrm{TOT}=0.43\arcsec$ in reproducing the observed positions of the 149 multiple images used as constraints in the lens model (see \Sec \ref{sec:MassParametrization}). In \Fig\ref{fig:RMS}, we show for each multiple image the displacement $\boldsymbol{\Delta}_{i}$ between its observed and model predicted positions. To quantify the precision of the lens model around the three main massive structures within the considered area (see \Fig \ref{fig:TotClusterCM}), we adopt different colors to distinguish the multiple images forming around the main cluster BCGs (red), the two bright galaxies G1 and G2 (green), and the cluster galaxy G3 (blue). We note that the same color scheme is adopted in \Fig \ref{fig:TotClusterImage} to indicate the observed positions of the multiple images. 
The positions of the 117 multiple images forming around the main cluster BCGs are reproduced with a precision of $\boldsymbol{\Delta}_{rms}^\mathrm{BCGs}=0.46\arcsec$. This is a remarkable result considering that this value is just $0.09\arcsec$ larger than that obtained using the previous model by \citetalias{Bergamini2022}, which included 27 fewer multiple images in the same region. 
An even higher precision of $\boldsymbol{\Delta}_{rms}^\mathrm{G1,G2}=0.20\arcsec$ is found in reproducing the observed positions of the 29 multiple images forming around G1 and G2. In contrast, the \EM\ precision decreases to $\boldsymbol{\Delta}_{rms}^\mathrm{G3}=0.73\arcsec$ around galaxy G3. Here, only three non-spectroscopic images (700.a, 700.b, and 700.c) of a single background galaxy are included in the lens model. We notice that the deflecting contribution of the massive clump surrounding G3 is mainly driven by its influence on the many multiple images observed relatively far from its center, i.e. in the cluster core. For this reason, the optimized values of the parameters of this clump might not be able to reproduce as well as for the other sources the three multiple images appearing close to its center.

A further confirmation of our previous results comes from the cumulative projected total mass profiles, presented in \Fig \ref{fig:mass}. These profiles and the associated statistical errors are computed considering 500 realizations obtained by randomly extracting parameter samples from the \EM\ MCMC chains. The black and red solid lines correspond to the median mass profiles, while the colored shaded bands are limited by the 16th and 84th percentiles.
Thanks to the large number of spectroscopically confirmed multiple images observed around the main cluster core, the total mass distribution of the cluster is reconstructed robustly in this region (top-left panel of the figure). This results in a cumulative projected total mass profile measured with very small statistical errors (black band in the figure). By comparing the results from this \EM\ with those by \citetalias{Bergamini2022}, we find a remarkably good agreement. In particular, the cumulative projected total mass profiles of the two models differ by less than 3\% close to the BCG-N core and less than 1\% at distances between 20\,kpc and 310\,kpc. 
Unsurprisingly, larger differences are found in the regions surrounding the three external clumps. On the top-right and bottom-left panels of \Fig \ref{fig:mass}, we plot the cumulative projected total mass profiles as a function of the distance from the G1 and G2 galaxies, respectively. These profiles demonstrate that the inclusion in the \EM\ of the multiple images identified with JWST/NIRCam imaging around G1 and G2 reduces significantly the statistical uncertainty in the total mass estimates. This is evidenced by the smaller extension of the black bands with respect to the red ones. While the total mass profile around G1 is compatible with the previous result by \citetalias{Bergamini2022}, a clear difference between the total mass profiles is found in the vicinity of G2. The newly identified multiple images around G1 and G2 -- especially the 13 lensed substructures of the spectroscopically confirmed spiral galaxy at $z=2.584$ (see \Fig \ref{fig:spec}) -- are thus key to precisely reconstructing the total mass distribution in this region. Similarly, we remark a large discrepancy between the total mass profiles obtained from the \EM\ and the \citetalias{Bergamini2022} model, as displayed in the bottom-right panel of \Fig \ref{fig:mass}. In this case, the three non-spectroscopic multiple images, from a single background source, lying close to the galaxy G3 can only poorly constrain the total mass distribution of the G3 halo. This is reflected into the black error band which is more extended than those in the other panels.
We remark that the new subhalo scaling relation is consistent over more than 5 magnitudes with that measured by \citetalias{Bergamini2022} with MUSE spectroscopy, within the $1\sigma$ statistical uncertainties (see \T \ref{table:inout_lensing} and their Figure 4). 
This highlights the importance of an independent determination of the scaling relation in reducing inherent model degeneracies between the cluster- and subhalo mass components.

On the left panel of \Fig \ref{fig:magnificatio_maps}, we show the relative difference between the median magnification values derived from the \EM\ ($\overline{\mu}_\mathrm{EX}$) and \citetalias{Bergamini2022} model ($\overline{\mu}_\mathrm{B23}$). The median magnification maps are obtained by considering 500 random realizations of the lens models and computing the median magnification value in each pixel. The right panel of \Fig \ref{fig:magnificatio_maps} shows instead the ratio of the relative errors on the predicted magnification values for the two models. In this case, the absolute errors $\Delta\mu_\mathrm{EM}$ and $\Delta\mu_\mathrm{B22}$ are computed from the 16th and 84th percentiles of the magnification distributions in each pixel. 

In the region surrounding the main cluster, the \EM\ and \citetalias{Bergamini2022} models predict very similar magnification values and associated errors (see the bottom-left region in the maps). This result is corroborated by the similarity of the \citetalias{Bergamini2022} (magenta) and \EM\ (white) critical lines displayed in \Fig \ref{fig:TotClusterImage} for a source at $z_s=6$. Within the main cluster, non-negligible differences between the predicted magnification are observed only at distances beyond $\sim28\arcsec$ North-West from the BCG-N, a region deprived of observed multiple images. The main cluster critical line predicted by the best-fit \EM\ becomes resonant with the secondary critical lines produced by two bright galaxies. 
The consistency between the \citetalias{Bergamini2022} and the \EM\ models in the main cluster region demonstrates that the modeling parameterization from \citetalias{Bergamini2022} is well suited to robustly characterize its total mass distribution.
As expected, closer to the external clumps surrounding galaxies G1, G2, and G3 (in the top and top-right regions of \Fig \ref{fig:magnificatio_maps}) the differences between the \EM\ and \citetalias{Bergamini2022} models are significant. In particular, the magnification maps reach relative differences of more than $\pm100\%$ in addition to the significant discrepancy on the shape of the critical lines. In fact, the critical lines obtained by the new \EM\ are significantly more centrally concentrated with respect to the large critical lines produced by the simplistic SIS mass distributions used by \citetalias{Bergamini2022} to parameterize the external clumps, illustrating the importance of adding strong lensing constraints in those regions.  
The right panel of \Fig \ref{fig:magnificatio_maps} demonstrates that the \EM\ model is characterized by similar relative errors to \citetalias{Bergamini2022} in the main cluster region in predicting the magnification values, whereas significantly smaller relative errors are found around the external clumps compared to \citetalias{Bergamini2022}. Particular attention has to be paid to the red ring-like and dark blue areas centered on the external clumps. These regions wrap the critical lines of the \citetalias{Bergamini2022} model and \EM\ computed for a source at $z_s=3$ (corresponding to the redshift of the magnification maps). 
We note that the magnification values presented by \citet{Castellano2022} for the $z\sim 10$ galaxies, based on \citetalias{Bergamini2022}, all in the low$-\mu$ regime ($\mu<3.8$), are well in agreement with the values computed with the new lens model.   

An additional way to demonstrate the robustness of the lens model is to compare the model-predicted redshift values of the non-spectroscopic systems with the photometric redshift estimates for their multiple images. 
In \Fig \ref{fig:zlens}, we show a comparison between the probability density distributions of the redshift values obtained from the \EM\ lens model (gray histogram) and the photometric redshift estimated by exploiting the JWST and HST multi-band data (see \Sec \ref{sec:imagingdata}) as the colored histograms, for the six non-spectroscopic multiple image systems included in the lens model. The remaining two non-spectroscopic systems (i.e., Sys-301 and Sys-200) used in the model do not have reliable photometric redshift measurements for any of their images, due to their faintness and to light contamination from nearby objects and, for that reason, are not included in the figure. 
The photometric Sys-7, Sys-28, and Sys-53 are located in the main cluster field (around the cluster BCGs) and have three multiple images each, corresponding to the nine red circles in \Fig \ref{fig:TotClusterImage}. From the JWST and HST multi-band observations, we obtain consistent photometric redshift estimates for two out of the three multiple images of Sys-7 and Sys-28 (i.e., 7a, 7c, 28a, and 28b). While for Sys-28 the photometric redshifts are in very good agreement with the \EM\ predicted redshift, we find that the model-predicted redshift of Sys-7 is slightly underestimated compared to the photometric estimates. 
Regarding Sys-53, the photometric redshift probability distribution for image 53b peaks at $z^{peak}_{ph}(53b)=1.575$, that is well below the photometric redshift values inferred for the other two multiple images, 53a and 53c (yellow and green histograms in the plot, respectively) and the model-predicted redshift. The likely reason for this discrepancy is the light contamination produced by a cluster galaxy (with $m_\mathrm{F160W}=22.05$) residing at a projected distance of just $0.9\arcsec$ from the image 53b. Moreover, \cite{Furtak2022_2} have measured a photometric redshift of $z_{ph}(\rm {Sys\mbox{-}53})\sim7.7$, in very good agreement with our model-predicted redshift and with the photometric redshifts of the images 53a and 53c from this work.
Sys-601 and Sys-602 both reside around the bright galaxy G2 and count
three and four multiple images, respectively (these correspond to the seven green circles around the galaxy G2 in \Fig \ref{fig:TotClusterImage}, see also the dashed white squares in \Fig \ref{fig:spec}). The multi-band photometric data allow for the estimate of the photometric redshifts for all the multiple images of the systems. These are found to be compatible among each other and with the lens model-predicted redshift probability density distributions. We note that we obtain identical redshift values for the two systems. Considering that the Sys-601 and Sys-602 sources are just $2.1\arcsec$ apart (corresponding to 17.5\,kpc at $z=2.3$), we can argue that these are likely two galaxies belonging to the same gravitationally bound system. We also note that the positions of the images of the Sys-601 and Sys-602 are extremely well reproduced by the \EM\ with values of $\Delta_{rms}^\mathrm{Sys\mbox{-}601}=0.28\arcsec$ and $\Delta_{rms}^\mathrm{Sys\mbox{-}601}=0.22\arcsec$.
Finally, Sys-700 consists of three multiple images forming around the galaxy G3 (blue circles in \Fig \ref{fig:TotClusterImage}). Also in this case, the lens model-predicted redshift is in very good agreement with the photometric redshift estimated for the single image 700.1c.

\section{Conclusions} \label{sec:conclusions}

We have presented a new high-precision strong lensing model of the galaxy cluster Abell 2744 ($z=0.3072$) based on the deep, wide-field JWST/NIRCam imaging \citep{Treu2022, Bezanson2022,Paris2023} and new MUSE DDT observations, together with extensive archival spectro-photometric datasets including two recent spectroscopic redshifts with JWST/NIRISS and JWST/NIRSpec \citep{Vanzella2022, Lin2023, Roberts-Borsani2022}. The model is constrained by 149 multiple images (121 of which are spectroscopically confirmed) from 50 background sources, spanning a very wide redshift range between $1.03$ and $9.76$. The multiple image catalog compiled for this cluster is the largest to date (see \Fig \ref{fig:images}), and represents an increase of about $66\%$ with respect to our previous, HST-based lens model (\citetalias{Bergamini2022}).
Specifically, we have included 27 additional multiple images in the main cluster core (18 of which are spectroscopically confirmed) thanks to a re-analysis of faint multiple images in the archival MUSE and the JWST/NIRISS data.
In addition, the deep, high-resolution JWST/NIRCam observations have proved to be key in the identification of a large number of multiple image systems around the external clumps, in particular those associated with the bright galaxies G1 and G2 (see also \citetalias{Furtak2022}), where we have considered 29 multiple images from 10 background sources. By analyzing the new data from the MUSE DDT program, we have presented the first spectroscopic redshift of a multiple image system in the G1-G2 external clump, a face-on spiral galaxy at $z=2.584$.

The model includes 177 cluster member galaxies, 163 of which are spectroscopically confirmed (24 of them are newly confirmed with the new MUSE data), and 14 are securely identified as cluster members with a CNN deep-learning technique \citep{Angora_2020}.
The new subhalo scaling relation is consistent, within the $1\sigma$ statistical uncertainties, with the measured one shown by \citetalias{Bergamini2022}, demonstrating the accurate total mass modeling of the lens cluster.

The total root-mean-square separation between the observed and model-predicted positions of the 149 multiple images is $\boldsymbol{\Delta}_{rms}^\mathrm{TOT}=0.43\arcsec$ on the image plane (see \Fig \ref{fig:RMS}). The achieved precision is similar to that obtained by \citetalias{Bergamini2022} ($\boldsymbol{\Delta}_{rms}^\mathrm{TOT}=0.37\arcsec$), despite the significantly larger number of multiple images considered in this work. 
We note that the precision of our model is significantly better than that of the JWST-based lens model presented by \citetalias{Furtak2022}, who quote a $\boldsymbol{\Delta}_{rms}^\mathrm{TOT}=0.66\arcsec$ on the lens plane.
The newly included multiple image systems around the external clumps represent a leap forward in the lens modeling of Abell 2744 in terms of precision and accuracy.
While the cumulative projected total mass profiles and the magnification maps obtained from our new model are highly consistent with those obtained by \citetalias{Bergamini2022} in the main cluster core, these quantities are robustly reconstructed at larger radii with significantly smaller statistical uncertainties (see Figures \ref{fig:mass} and \ref{fig:magnificatio_maps}) thanks to the newly included constraints. 
The robustness of the lens model is further demonstrated by the consistency between the model-predicted and photometric redshift estimates of 6 non-spectroscopic multiple image systems included in the model (see Figure \ref{fig:zlens}).  

Our new lens model holds a fundamental role in the study of the lensed high-redshift sources that are being observed by the JWST in this cluster field \citep[e.g.,][]{Castellano2022, Prieto-Lyon2022, Morishita2022}, in particular in the high-magnification regime \citep[e.g.,][]{Roberts-Borsani2022, Vanzella2022}. Further improvements will be possible when additional multiple image systems will be spectroscopically confirmed with on-going JWST spectroscopic observations.

The lens model presented in this work will be made publicly available with the publication of this paper through our recently developed Strong Lensing Online Tool, \SLOT\ (\citetalias{Bergamini2022}), allowing researchers to take full advantage of the predictive and statistical results of our lens model through a user-friendly graphical interface. The complete catalog of secure redshift measurements (i.e., QF $\ge$ 2) from the new VLT/MUSE DDT observations (see \Sec \ref{sec:dataSpec}) will also be released upon publication.

\begin{acknowledgments}
Support for program JWST-ERS-1324 was provided by NASA through a grant from the Space Telescope Science Institute, which is operated by the Association of Universities for Research in Astronomy, Inc., under NASA contract NAS 5-03127.
The Hubble Frontier Field program (HFF) and the Beyond Ultra-deep Frontier Fields And Legacy Observations (BUFFALO) are based on the data made with the NASA/ESA {\it Hubble Space Telescope}. The Space Telescope Science Institute is operated by the Association of Universities for Research in Astronomy, Inc., under NASA contract NAS 5-26555. ACS was developed under NASA Contract NAS 5-32864. Based also on observations collected at the European Southern Observatory for Astronomical research in the Southern Hemisphere under ESO programmes with IDs 094.A-0115 (PI: Richard).
We acknowledge financial support through grants PRIN-MIUR 2015W7KAWC, 2017WSCC32, and 2020SKSTHZ. AA has received funding from the European Union’s Horizon 2020 research and innovation programme under the Marie Skłodowska-Curie grant agreement No 101024195 — ROSEAU. GBC thanks the Max Planck Society for support through the Max Planck Research Group for S. H. Suyu and the academic support from the German Centre for Cosmological Lensing. MM acknowledges support from the Italian Space Agency (ASI) through contract ``Euclid - Phase D". We acknowledge funding from the INAF ``main-stream'' grants 1.05.01.86.20 and  1.05.01.86.31. We acknowledge support from the INAF Large Grant 2022 “Extragalactic Surveys with JWST”  (PI: Pentericci).
This research is supported in part by the Australian Research Council Centre of Excellence for All Sky Astrophysics in 3 Dimensions (ASTRO 3D), through project number CE170100013.
The GLASS-JWST data are associated with DOI 10.17909/mrt6-wm89 (NIRCam) and 10.17909/9a2g-sj78 (NIRSpec).
The UNCOVER NIRcam data are associated with DOI 10.17909/zn4s-0243. The JWST data from the DDT program 2756 are associated with DOI 10.17909/te6f-cg91.
\end{acknowledgments}

%

\vspace{5mm}
\facilities{JWST(NIRCam), JWST(NIRISS), VLT(MUSE), HST(ACS), HST(WFC3)}





\clearpage
\appendix

\section{Catalog of multiple images in Abell 2744}
We present here the complete catalog of 
multiple images included in the new lens model in the JWST WCS (see \Sec \ref{sec:imagingdata}). For consistency, we also list the multiple images presented in the \citetalias{Bergamini2022} HST-based lens model (marked with a dagger symbol, $\dagger$) that were anchored to the HFF WCS. The multiple images identified with a star symbol, $\star$, indicate those newly spectrocopically confirmed with the central MUSE datacube, while those with an asterisk, $\ast$,  have been spectroscopically confirmed with JWST/NIRISS \citep{Vanzella2022, Lin2023} and JWST/NIRSpec \citep{Roberts-Borsani2022}.
Further details can be found in \citetalias{Bergamini2022} and in \Sec \ref{sec:multiple_images}.

\begin{longtable}{ccccccc}
\centering
\label{tab:multiple_images}

ID & R.A. & Dec. & QP & z & QF & Location \\
 & [deg.] & [deg.] &  &  & &  \\
\hline

1.1a$\dagger$ & $3.597551$ & $-30.403906$ & 1 & 1.688 & 3 & BGCs \\
1.1b$\dagger$ & $3.595952$ & $-30.406787$ & 1 & 1.688 & 2 & BGCs \\
1.1c$\dagger$ & $3.586210$ & $-30.409963$ & 1 & 1.688 & 3 & BGCs \\
1.2a$\dagger$ & $3.597062$ & $-30.404703$ & 1 & 1.688 & 2 & BGCs \\
1.2b$\dagger$ & $3.596384$ & $-30.406123$ & 1 & 1.688 & 3 & BGCs \\
1.2c$\dagger$ & $3.585732$ & $-30.410077$ & 1 & 1.688 & 3 & BGCs \\
1.3a$\dagger$ & $3.597746$ & $-30.403511$ & 2 & 1.688 & 3 & BGCs \\
1.3b$\dagger$ & $3.595516$ & $-30.407178$ & 2 & 1.688 & 3 & BGCs \\
1.3c$\dagger$ & $3.586444$ & $-30.409849$ & 2 & 1.688 & 3 & BGCs \\
1.4a$\dagger$ & $3.598073$ & $-30.403961$ & 1 & 1.688 & 1 & BGCs \\
1.4b$\dagger$ & $3.595710$ & $-30.407526$ & 1 & 1.688 & 1 & BGCs \\
1.4c$\dagger$ & $3.587368$ & $-30.410130$ & 1 & 1.688 & 1 & BGCs \\
2.1a$\dagger$ & $3.583251$ & $-30.403317$ & 1 & 1.887 & 3 & BGCs \\
2.1b$\dagger$ & $3.597281$ & $-30.396694$ & 1 & 1.887 & 3 & BGCs \\
2.1c$\dagger$ & $3.585356$ & $-30.399856$ & 1 & 1.887 & 3 & BGCs \\
2.1d$\dagger$ & $3.586400$ & $-30.402106$ & 1 & 1.887 & 3 & BGCs \\
2.2a$\dagger$ & $3.583015$ & $-30.403167$ & 1 & 1.887 & 3 & BGCs \\
2.2b$\dagger$ & $3.597130$ & $-30.396620$ & 1 & 1.887 & 3 & BGCs \\
2.2c$\dagger$ & $3.585122$ & $-30.399647$ & 1 & 1.887 & 3 & BGCs \\
2.2d$\dagger$ & $3.586425$ & $-30.401849$ & 1 & 1.887 & 3 & BGCs \\
2.3a$\dagger$ & $3.582980$ & $-30.403028$ & 1 & 1.887 & 3 & BGCs \\
2.3b$\dagger$ & $3.597087$ & $-30.396561$ & 1 & 1.887 & 3 & BGCs \\
2.3c$\dagger$ & $3.585004$ & $-30.399603$ & 1 & 1.887 & 3 & BGCs \\
2.3d$\dagger$ & $3.586381$ & $-30.401744$ & 1 & 1.887 & 3 & BGCs \\
2.4a$\dagger$ & $3.582905$ & $-30.402908$ & 1 & 1.887 & 3 & BGCs \\
2.4b$\dagger$ & $3.597044$ & $-30.396509$ & 1 & 1.887 & 3 & BGCs \\
2.4c$\dagger$ & $3.584902$ & $-30.399559$ & 1 & 1.887 & 3 & BGCs \\
2.4d$\dagger$ & $3.586340$ & $-30.401611$ & 1 & 1.887 & 3 & BGCs \\
2.5a$\dagger$ & $3.582814$ & $-30.402771$ & 2 & 1.887 & 3 & BGCs \\
2.5b$\dagger$ & $3.596982$ & $-30.396451$ & 2 & 1.887 & 3 & BGCs \\
2.5c$\dagger$ & $3.584807$ & $-30.399495$ & 2 & 1.887 & 3 & BGCs \\
2.5d$\dagger$ & $3.586309$ & $-30.401469$ & 1 & 1.887 & 3 & BGCs \\
2.6a$\dagger$ & $3.582758$ & $-30.402661$ & 2 & 1.887 & 3 & BGCs \\
2.6b$\dagger$ & $3.596934$ & $-30.396403$ & 2 & 1.887 & 3 & BGCs \\
2.6c$\dagger$ & $3.584710$ & $-30.399442$ & 2 & 1.887 & 3 & BGCs \\
2.6d$\dagger$ & $3.586272$ & $-30.401331$ & 2 & 1.887 & 3 & BGCs \\
2.7a$\dagger$ & $3.582516$ & $-30.402291$ & 2 & 1.887 & 3 & BGCs \\
2.7b$\dagger$ & $3.596726$ & $-30.396273$ & 2 & 1.887 & 3 & BGCs \\
2.7c$\dagger$ & $3.584460$ & $-30.399268$ & 2 & 1.887 & 3 & BGCs \\
2.7d$\dagger$ & $3.586224$ & $-30.400848$ & 2 & 1.887 & 3 & BGCs \\
3.1a$\dagger$ & $3.589365$ & $-30.393837$ & 1 & 3.98 & 3 & BGCs \\
3.1b$\dagger$ & $3.588797$ & $-30.393777$ & 1 & 3.98 & 3 & BGCs \\
3.1c$\ast$ & $3.576622$ & $-30.401792$ & 3 & 3.98 & 1 & BGCs \\
3.2a$\dagger$ & $3.589212$ & $-30.393824$ & 1 & 3.98 & 3 & BGCs \\
3.2b$\dagger$ & $3.588955$ & $-30.393795$ & 1 & 3.98 & 3 & BGCs \\
3.3a$\dagger$ & $3.589477$ & $-30.393850$ & 1 & 3.98 & 3 & BGCs \\
3.3b$\dagger$ & $3.588629$ & $-30.393765$ & 1 & 3.98 & 3 & BGCs \\
4.1a$\dagger$ & $3.592115$ & $-30.402640$ & 1 & 3.577 & 3 & BGCs \\
4.1b$\dagger$ & $3.595664$ & $-30.401614$ & 1 & 3.577 & 3 & BGCs \\
4.1c$\dagger$ & $3.580437$ & $-30.408922$ & 2 & 3.577 & 3 & BGCs \\
4.1e$\dagger$ & $3.593640$ & $-30.405097$ & 3 & 3.577 & 3 & BGCs \\
4.2a$\dagger$ & $3.592087$ & $-30.402507$ & 1 & 3.577 & 3 & BGCs \\
4.2b$\dagger$ & $3.595557$ & $-30.401499$ & 1 & 3.577 & 3 & BGCs \\
6.1a$\dagger$ & $3.598532$ & $-30.401773$ & 1 & 2.017 & 3 & BGCs \\
6.1b$\dagger$ & $3.594046$ & $-30.407978$ & 1 & 2.017 & 3 & BGCs \\
6.1c$\dagger$ & $3.586418$ & $-30.409340$ & 1 & 2.017 & 3 & BGCs \\
8.1a$\dagger$ & $3.589702$ & $-30.394316$ & 2 & 3.977 & 2 & BGCs \\
8.1b$\dagger$ & $3.588821$ & $-30.394182$ & 2 & 3.977 & 2 & BGCs \\
10a$\star$ & $3.588386$ & $-30.405852$ & 1 & 2.657 & 2 & BGCs \\
10b$\star$ & $3.587369$ & $-30.406455$ & 1 & 2.657 & 3 & BGCs \\
18.1a$\dagger$ & $3.576110$ & $-30.404448$ & 1 & 5.662 & 3 & BGCs \\
18.1b$\dagger$ & $3.588367$ & $-30.395610$ & 1 & 5.662 & 3 & BGCs \\
18.1c$\dagger$ & $3.590720$ & $-30.395523$ & 1 & 5.662 & 3 & BGCs \\
22.1a$\dagger$ & $3.587908$ & $-30.411586$ & 2 & 5.284 & 3 & BGCs \\
22.1b$\dagger$ & $3.600039$ & $-30.404396$ & 2 & 5.284 & 3 & BGCs \\
22.1c$\dagger$ & $3.596580$ & $-30.408968$ & 2 & 5.284 & 3 & BGCs \\
24a$\star$ & $3.595886$ & $-30.404448$ & 1 & 1.044 & 3 & BGCs \\
24b$\star$ & $3.595112$ & $-30.405881$ & 1 & 1.044 & 9 & BGCs \\
24c$\star$ & $3.587311$ & $-30.409064$ & 1 & 1.044 & 1 & BGCs \\
26.1a$\dagger$ & $3.593885$ & $-30.409703$ & 1 & 3.054 & 3 & BGCs \\
26.1b$\dagger$ & $3.590339$ & $-30.410553$ & 1 & 3.054 & 3 & BGCs \\
26.1c$\dagger$ & $3.600103$ & $-30.402920$ & 2 & 3.054 & 9 & BGCs \\
26.2a$\dagger$ & $3.593979$ & $-30.409677$ & 1 & 3.054 & 3 & BGCs \\
26.2b$\dagger$ & $3.590257$ & $-30.410588$ & 1 & 3.054 & 3 & BGCs \\
26.3a$\dagger$ & $3.594018$ & $-30.409583$ & 3 & 3.054 & 2 & BGCs \\
26.3b$\dagger$ & $3.589954$ & $-30.410571$ & 3 & 3.054 & 2 & BGCs \\
30a$\star$ & $3.591005$ & $-30.397415$ & 1 & 1.026 & 2 & BGCs \\
30b$\star$ & $3.586672$ & $-30.398157$ & 1 & 1.026 & 9 & BGCs \\
30c$\star$ & $3.581915$ & $-30.401675$ & 1 & 1.026 & 1 & BGCs \\
31a$\star$ & $3.585930$ & $-30.403132$ & 1 & 4.757 & 3 & BGCs \\
31b$\star$ & $3.583708$ & $-30.404073$ & 1 & 4.757 & 3 & BGCs \\
33.1a$\dagger$ & $3.584699$ & $-30.403125$ & 1 & 5.726 & 3 & BGCs \\
33.1b$\dagger$ & $3.584383$ & $-30.403371$ & 1 & 5.726 & 3 & BGCs \\
34.1a$\dagger$ & $3.593414$ & $-30.410812$ & 1 & 3.784 & 2 & BGCs \\
34.1b$\dagger$ & $3.593798$ & $-30.410692$ & 1 & 3.784 & 3 & BGCs \\
34.1c$\dagger$ & $3.600574$ & $-30.404413$ & 1 & 3.784 & 2 & BGCs \\
41a$\star$ & $3.599143$ & $-30.399557$ & 1 & 4.91 & 9 & BGCs \\
41b$\star$ & $3.593505$ & $-30.407730$ & 1 & 4.91 & 9 & BGCs \\
41c$\star$ & $3.583441$ & $-30.408475$ & 1 & 4.91 & 9 & BGCs \\
41d$\star$ & $3.590685$ & $-30.404541$ & 3 & 4.91 & 2 & BGCs \\
42.1a$\dagger$ & $3.597304$ & $-30.400586$ & 1 & 3.692 & 3 & BGCs \\
42.1b$\dagger$ & $3.590944$ & $-30.403231$ & 1 & 3.692 & 3 & BGCs \\
42.1c$\dagger$ & $3.581575$ & $-30.408608$ & 1 & 3.692 & 3 & BGCs \\
42.1d$\dagger$ & $3.594233$ & $-30.406367$ & 1 & 3.692 & 3 & BGCs \\
42.1e$\dagger$ & $3.592403$ & $-30.405176$ & 3 & 3.692 & 1 & BGCs \\
61.1a$\dagger$ & $3.595511$ & $-30.403465$ & 1 & 2.951 & 1 & BGCs \\
61.1b$\dagger$ & $3.595127$ & $-30.404451$ & 1 & 2.951 & 3 & BGCs \\
62.1a$\dagger$ & $3.591315$ & $-30.398623$ & 3 & 4.194 & 3 & BGCs \\
62.1b$\dagger$ & $3.590571$ & $-30.398898$ & 3 & 4.194 & 3 & BGCs \\
63.1a$\dagger$ & $3.582199$ & $-30.407119$ & 1 & 5.662 & 3 & BGCs \\
63.1b$\dagger$ & $3.592823$ & $-30.407011$ & 2 & 5.662 & 3 & BGCs \\
63.1c$\dagger$ & $3.589141$ & $-30.403406$ & 2 & 5.662 & 3 & BGCs \\
63.1d$\dagger$ & $3.598822$ & $-30.398255$ & 2 & 5.662 & 3 & BGCs \\
64.1a$\dagger$ & $3.581190$ & $-30.398712$ & 2 & 3.409 & 3 & BGCs \\
64.1c$\dagger$ & $3.596412$ & $-30.394247$ & 2 & 3.409 & 3 & BGCs \\
JD1A$\ast$ & $3.592492$ & $-30.401458$ & 1 & 9.756 & 2 & BGCs \\
JD1B$\ast$ & $3.595004$ & $-30.400730$ & 1 & 9.756 & 2 & BGCs \\
JD1C$\ast$ & $3.577512$ & $-30.408667$ & 1 & 9.756 & 2 & BGCs \\
600.1a & $3.543063$ & $-30.368102$ & 1 & 2.58 & 2 & G1-G2 \\
600.1b & $3.540339$ & $-30.372503$ & 1 & 2.58 & 2 & G1-G2 \\
600.1c & $3.540449$ & $-30.375070$ & 1 & 2.58 & 2 & G1-G2 \\
600.1d & $3.542075$ & $-30.372619$ & 1 & 2.58 & 2 & G1-G2 \\
600.3a & $3.542649$ & $-30.368409$ & 1 & 2.58 & 2 & G1-G2 \\
600.3b & $3.540129$ & $-30.372309$ & 1 & 2.58 & 2 & G1-G2 \\
600.3c & $3.540147$ & $-30.375323$ & 1 & 2.58 & 2 & G1-G2 \\
600.4a & $3.543199$ & $-30.367893$ & 1 & 2.58 & 2 & G1-G2 \\
600.4b & $3.540398$ & $-30.372637$ & 1 & 2.58 & 2 & G1-G2 \\
600.4c & $3.540540$ & $-30.374834$ & 1 & 2.58 & 2 & G1-G2 \\
600.5a & $3.542933$ & $-30.367952$ & 1 & 2.58 & 2 & G1-G2 \\
600.5b & $3.540186$ & $-30.372640$ & 1 & 2.58 & 2 & G1-G2 \\
600.5c & $3.540297$ & $-30.374845$ & 1 & 2.58 & 2 & G1-G2 \\
\hline
7a & $3.598254$ & $-30.402310$ & 1 & - & - & BGCs \\
7b & $3.595216$ & $-30.407384$ & 1 & - & - & BGCs \\
7c & $3.584596$ & $-30.409797$ & 1 & - & - & BGCs \\
28a & $3.580439$ & $-30.405018$ & 1 & - & - & BGCs \\
28b & $3.597824$ & $-30.395939$ & 1 & - & - & BGCs \\
28c & $3.585308$ & $-30.397926$ & 1 & - & - & BGCs \\
53a & $3.579837$ & $-30.401562$ & 1 & - & - & BGCs \\
53b & $3.583540$ & $-30.396673$ & 1 & - & - & BGCs \\
53c & $3.597214$ & $-30.394337$ & 1 & - & - & BGCs \\
A200.1a & $3.556366$ & $-30.376714$ & 1 & - & - & G1-G2 \\
A200.1b & $3.556536$ & $-30.375809$ & 1 & - & - & G1-G2 \\
B200.2a & $3.556331$ & $-30.376780$ & 1 & - & - & G1-G2 \\
B200.2b & $3.556531$ & $-30.375736$ & 1 & - & - & G1-G2 \\
C200.3a & $3.556468$ & $-30.376380$ & 1 & - & - & G1-G2 \\
C200.3b & $3.556511$ & $-30.376159$ & 1 & - & - & G1-G2 \\
301.1a & $3.554637$ & $-30.373839$ & 1 & - & - & G1-G2 \\
301.1b & $3.553188$ & $-30.375998$ & 1 & - & - & G1-G2 \\
301.1c & $3.552300$ & $-30.378605$ & 1 & - & - & G1-G2 \\
601.1a & $3.544279$ & $-30.368064$ & 1 & - & - & G1-G2 \\
601.1b & $3.542829$ & $-30.373490$ & 1 & - & - & G1-G2 \\
601.1c & $3.542022$ & $-30.374625$ & 1 & - & - & G1-G2 \\
602.1a & $3.543093$ & $-30.369106$ & 1 & - & - & G1-G2 \\
602.1b & $3.540729$ & $-30.375868$ & 1 & - & - & G1-G2 \\
602.1c & $3.542623$ & $-30.372453$ & 1 & - & - & G1-G2 \\
602.1d & $3.540913$ & $-30.372220$ & 1 & - & - & G1-G2 \\
700.1a & $3.579703$ & $-30.357717$ & 1 & - & - & G3 \\
700.1b & $3.579165$ & $-30.357833$ & 1 & - & - & G3 \\
700.1c & $3.574908$ & $-30.356075$ & 1 & - & - & G3 \\
\hline
\\
\caption{List of the multiple images used as constraints in the \texttt{Extended Model} we are presenting. The multiple images of the same background source share the same numerical part of the ID (first column). In the fourth column, we quote the redshifts of the sources. 
The `positional quality flag', QP, quantifies the precision in determining the on-sky position of the multiple images. QP=1 corresponds to images with a compact HST or JWST emission. The positional error assumed on these images in the lens model is the lowest. QP=2 stands for a diffuse or elongated HST or JWST emission. An intermediate positional error is associated to these images. Images with QP=3 are only detected in the MUSE datacube or are characterized by a very diffuse emission. These have the largest positional errors. In the last column, we report the region of the cluster where the multiple images are located. These regions are identified using the name of the brightest galaxies they contain (see \Fig \ref{fig:TotClusterImage}).}
\end{longtable}

\section{Redshift catalog}
In this section, we present an extract of the catalog of 
all the sources for which we measure a secure redshift value (QF~$\ge$~2) from the new MUSE data obtained during the ESO DDT program 109.24EZ.001 (see \Sec \ref{sec:dataSpec} and \Fig \ref{fig:TotClusterCM}). The full catalog will be released upon publication. 

\begin{longtable}{ccccc}
\label{tab:ddtcatalog}
ID & R.A. & Dec. & z & QF \\
 & [deg.] & [deg.] &  &  \\
\hline

70000123 & $3.520965$ & $-30.372848$ & 0.000 & 3 \\
70004495 & $3.499187$ & $-30.323966$ & 0.000 & 3 \\
70005368 & $3.482608$ & $-30.318157$ & 0.000 & 3 \\
70001070 & $3.543078$ & $-30.367872$ & 0.000 & 3 \\
70006188 & $3.479497$ & $-30.306998$ & 0.000 & 3 \\
... & ... & ... & ... & ... \\

\hline
\\
\caption{Extract of the catalog of the sources with a reliable redshift measurementfrom the VLT/MUSE data obtained during the ESO DDT program 109.24EZ.001. The values of the QF are described in \Sec \ref{sec:dataSpec}.}
\end{longtable}

\bibliography{bibliography}{}
\bibliographystyle{aasjournal}



\end{document}